\documentclass[english,11pt,aps,prd,nofootinbib,showkeys,preprint,floatfix]{revtex4-1}
\usepackage[utf8]{inputenc}
\usepackage[english]{babel}
\usepackage{amsmath,amssymb,amsfonts,amsthm}
\usepackage{graphicx}
\usepackage{bbold}
\usepackage[colorlinks=True]{hyperref}
\usepackage[usenames,dvipsnames,svgnames]{xcolor}
\newcommand{\AddrUdeA}{%
 Instituto de F\'\i sica, Universidad de Antioquia,\\
 Calle 70 No. 52-21, Medell\'in, Colombia. }

\begin{document}

\title{Multi-component secluded WIMP dark matter and Dirac neutrino masses with an extra Abelian gauge symmetry}

\author{Kimy Agudelo}\email{kimy.agudelo@udea.edu.co}
\affiliation{\AddrUdeA} 
\author{Diego Restrepo} \email{restrepo@udea.edu.co}
\affiliation{\AddrUdeA}
\author{Andr\'es Rivera} \email{afelipe.rivera@udea.edu.co}
\affiliation{\AddrUdeA}
\author{David Suarez} \email{david.suarezr@udea.edu.co}
\affiliation{\AddrUdeA}

\date{\small\today}


\begin{abstract}
Scenarios for secluded WIMP dark matter models have been extensively studied in simplified versions. This paper shows a complete UV realization of a secluded WIMP dark matter model with an extra Abelian gauge symmetry that includes two-component dark matter candidates, where the dark matter conversion process plays a significant role in determining the relic density in the Universe. The model contains two new unstable mediators: a dark Higgs and a dark photon. It generates Dirac neutrino masses and can be tested in future direct detection experiments of dark matter. The model is also compatible with cosmological and theoretical constraints, including the branching ratio of Standard model particles into invisible, Big Bang nucleosynthesis restrictions, and the number of relativistic degrees of freedom in the early Universe, even without kinetic mixing.
\end{abstract}

\maketitle

\section{Introduction}


Cosmology requires a heavy neutral stable particle to be a viable dark matter (DM) candidate, which could 
be a thermal relic like in the WIMP (weakly interactive dark matter)  
paradigm~\cite{Steigman:1984ac,Cushman:2013zza}. However, there are other 
possibilities in addition to WIMPs~\cite{Bertone:2016nfn,Feng:2022rxt}. 
For example: QCD Axions in the masses window $[10^{-12}-1] $ eV~\cite{Preskill:1982cy,Abbott:1982af,Dine:1982ah}, 
sterile neutrinos with masses in the keV range~\cite{Kopp:2021jlk}, primordial black holes (PBHs) 
that could form in the early Universe with around $30$ solar 
masses~\cite{Carr:2020gox,Bird:2016dcv,Barman:2024kfj}, etc. On the other hand, 
neutrino oscillation data need a beyond Standard Model (SM) plausible mechanism that generates 
either Majorana or Dirac neutrino masses; for example, the seesaw mechanisms (reviewed in~\cite{Chulia:2021jgv}) 
or some radiative mechanism (reviewed in~\cite{Cai:2017jrq}).

A popular paradigm of DM is the secluded WIMP~\cite{Pospelov:2007mp},
where the processes leading to the relic abundance of DM occur in the dark sector. 
In this scenario, DM candidates primarily annihilate into lighter mediators
that then decay into SM states.

When a local $U(1)_D$ secluded sector is added to the SM~\cite{Pospelov:2008zw},  
the kinetic mixing between the visible photon and the secluded $U(1)_D$ gauge boson 
(the dark photon) becomes possible \cite{Holdom:1985ag}, opening a decay channel 
for the dark photon into SM particles.
However, for vector-like Dirac Dark Matter (DM) in a $U(1)_D$ secluded sector, 
constraints on this kinetic mixing are often quite strong~\cite{Cline:2014dwa}, 
as reviewed in~\cite{Fabbrichesi:2020wbt} considering laboratory, astrophysical, 
cosmological observations, and DM physics.

%
%
%

Alternatively, the $U(1)_D$ symmetry can be broken via the
Brout-Englert-Higgs mechanism, similar to the SM. This generates a
dark Higgs~\cite{Bell:2016uhg} and provides a mechanism to generate mass for the
dark photon without relying on the Stückelberg 
mechanism~\cite{Stueckelberg:1938hvi}. 
Furthermore, we can use the Spontaneous Symmetry Breaking (SSB) to generate masses for 
chiral DM fermions through Yukawa interactions, which further relax the constraints which apply to the 
vector-like case~\cite{Cline:2014dwa,Hambye:2019dwd}. 
As a result, in this scenario, we have the emergence of two dark mediators, the dark photon and the dark Higgs 
and new Yukawa interactions, which open a rich phenomenology with some imprints in the relic density of DM.

To ensure anomaly cancellation in the secluded sector when all DM fermions acquire mass 
from the dark Higgs mechanism, at least five two-component chiral fermions 
are required~\cite{Costa:2019zzy}. 
Some of them can be massless and can play the role of the right-handed 
components of Dirac neutrinos, providing a new direct decay channel for the dark photon 
into neutrinos, which further relaxes the constraints on the kinetic mixing.
With charged right-handed neutrinos, global lepton number conservation is ensured and 
Dirac neutrino masses are forbidden at tree-level. 
When a Dirac neutrino mass operator is allowed at a non-renormalizable level, 
the specific sets of charges are fully determined, leading to a broad range of models.
In~\cite{Restrepo:2021kpq}, we identified approximately one thousand such models, 
considering sets of up to twelve two-component chiral fermions. 
Due to this large number of massive DM fermions, these models generally feature 
multi-component Dark Matter candidates.
Any one of these models can be realized through a radiative Dirac-seesaw with some of 
the heavy DM fermions, and one extended inert scalar sector~\cite{Wong:2020obo,Bernal:2021ezl}.
It is worth noticing that this mechanism to generate neutrino masses and mixings is compatible
with the non-observation of neutrinoless double beta decay.

These approximately one thousand models share common characteristics: they are secluded models featuring 
two mediators (the dark Higgs and dark photon) and contain multi-component Dark Matter candidates. 
Crucially, the dark photon possesses a novel decay mode into neutrinos, which can alleviate 
the stringent constraints typically associated with kinetic mixing. 

This work, for the first time, explores the phenomenology of this rich and compelling framework.
To isolate and highlight the impact of the dark photon's new decay mode into neutrinos when 
the inert scalar sector is decoupled, this exploration will be conducted using a specific model with 
the simplest possible inert scalar sector.
The specific model contains nine two-component chiral fermions 
which cancel out the anomalies in the secluded sector.
Six of these chiral fermions form three Dirac fermions, acquiring masses through the dark Higgs mechanism.
The first two Dirac fermions form a sector of two-flavors from which the lightest Dirac fermion 
is a first dark matter candidate. The third Dirac fermion constitutes a second DM candidate. 
Finally, the last three chiral fermions remain as massless right-handed neutrinos, 
and interact with the inert scalar sector and the two-flavor Dirac fermions, 
to generate one-loop Dirac neutrino masses.  
When the dark Higgs and the inert scalar sector are decoupled, the
DM candidates mainly annihilate into dark photons that later decay into neutrinos. 
Consequently, the proper value of the relic abundance of DM
is obtained.

We explain the generalities of the secluded models in
Sec.~\ref{sec:Secluded_gauge_models_with_chiral_fermions}. We describe
our model in Sec.~\ref{sec:model}. In Sec.~\ref{sec:DM}, we explain
the DM considerations within our model. In
Sec.~\ref{sec:phenomenology}, we present the phenomenology of the
model, the DM analysis, the numerical results, the status, and the
prospects for DM searches. We conclude in Sec.~\ref{sec:Conclusions}.

\section{Secluded gauge models with chiral fermions}
\label{sec:Secluded_gauge_models_with_chiral_fermions}

In a fundamental theory, the elementary fermions are expected to be
massless chiral fields. Their masses emerge through Yukawa terms,
which are allowed by the respective charges via some scalar that
triggers spontaneous symmetry breaking (SSB).  For a set of such
charges associated with the $N$ chiral fields under a $U(1)$ gauge
symmetry
\begin{align*}
  \boldsymbol{Z} = \left[Z_1,Z_2,\cdots,Z_N\right]\,,
\end{align*}
at least the linear anomaly (with one
$U(1)$ gauge boson and two gravitons on the external lines) and the cubic anomaly (with three $U(1)$ gauge
bosons on the external lines)
must be cancelled~\cite{Babu:2003is,Batra:2005rh,deGouvea:2015pea,Costa:2019zzy,Costa:2020dph},
\begin{align}
    \label{eq:Dcoond}
    \sum_{i=1}^{N} Z_i=&0\,,&
    \sum_{i=1}^{N} Z_i^3=&0\,.
\end{align}
In fact, in the standard model (SM), for each chiral generation, there
exists a solution to eq.~\eqref{eq:Dcoond} for the gauge Abelian
symmetry $U(1)_Y$.  Expressed in terms of 15 non-opposite-sign
integers, the solution~\footnote{There is factor $1/6$ between this
  convention and the more common one.}
\begin{align*}
  \boldsymbol{Y} = [\color{red}1, 1,\color{blue} 1, 1, \color{Green} 1, 1, 
  \color{red} -4, \color{blue} -4, \color{Green} \color{red}-4, 2,
  \color{blue} 2, \color{Green}2, {\color{black} -3,} {\color{black}-3,} 
  {\color{black}6}\color{black}],
\end{align*}
corresponds to the hypercharges of 15 left-handed Weyl massless
fermions.  The SSB is triggered by a Higgs, $\color{magenta}H$, of
hypercharge $\color{magenta}3$. This leads to Yukawa interactions
through the Dirac pairs
\begin{align*}
 u_\alpha=& \color{red}(1, -4), \color{blue}(1, -4), \color{Green}(1, -4),& 
 d_\alpha = & \color{red}(1, 2), \color{blue}(1, 2), \color{Green}(1, 2),& 
 e= & \color{black} (-3,6).
\end{align*}
Consequently, 14 of the chiral fermions acquire Dirac masses after the
SSB. Each set of quark Dirac fermions is degenerated due to the
additional color symmetry. A left-handed Weyl fermion, $\nu_L$, with
hypercharge $-3$, remains massless, while the remnant $Z_3$ symmetry
guarantees the stability of the lightest quark.

Similarly, the DM of the Universe would be part of a dark
sector that contains a self-consisting set of chiral fermions that
satisfy eq.~\eqref{eq:Dcoond} under the charges of a secluded gauge
Abelian symmetry, $U(1)_{D}$, where all the SM particles are neutral.
Following the SSB via a dark Higgs mechanism, the lightest massive
fermion could be a viable candidate for fermionic DM. This
work explores the general phenomenological features of a secluded dark
sector with chiral fermions and inert scalars where the dark photon, 
$Z^{\prime}$,
and the dark Higgs, $S$, act as mediators~\cite{Bell:2016uhg}.  Similarly to
the SM, the charges of the chiral fermions in the solution should not
allow a mass term larger than the scale of the SSB.  To relax the
constraints from the kinetic mixing~\cite{Holdom:1985ag}, those
models~\cite{delaVega:2023dmw} allow an additional annihilation
channel for the fermionic DM candidate into SM Dirac
neutrinos~\cite{Ma:2021szi}.  If the right-handed components of these
neutrinos, $\nu_R^\alpha$, 
couple directly to the dark photon, these Weyl fermions can
be identified with the remaining chiral massless fermions in the
solution after the SSB in the dark
sector~\cite{Ma:2021szi,Wong:2020obo,Bernal:2021ezl}.\footnote{In
  another class of secluded solutions, proposed to explain the reactor
  anti-neutrino anomaly~\cite{Mention:2011rk}, these fermions may also
  be interpreted as sterile
  neutrinos~\cite{Babu:2003is,Davoudiasl:2005ks,Heeck:2012bz,deGouvea:2015pea}. }

Models with a tree-level contribution to the Dirac neutrino masses 
are automatically forbidden. However, the effective Dirac neutrino mass 
operator can be allowed through the dark Higgs~\cite{Restrepo:2021kpq}
\begin{align}
\label{eq:neutrino-operator}
  \mathcal{L} =& w_{\alpha i} \left(
  \nu_{R\alpha}
  \right)^{\dagger} L_{i}H 
  S^{\delta}\,,&\delta = 1,2\,.
\end{align}
where $L_i$ are the SM lepton doublets, $H$ is the SM Higgs doublet, 
and $w_{\alpha i}$ are the Wilson coefficients. 
Notice that at least two right-handed neutrinos with secluded
degenerated charges are needed to explain the neutrino oscillation
data~\cite{Restrepo:2021kpq}.  

In the context of the $U(1)_D$ symmetry, we are interested in a self-consistent set of massless right-handed chiral fermions, with
$N'$ of them coupled to the dark Higgs,
and $\nu_R^{\alpha}$, 
$\alpha=1,\cdots,\alpha_{\text{max}}$ ($\alpha_{\text{max}}=2\text{ or }3$) with degenerate secluded charges $\nu$. The corresponding set of
$N = N'+\alpha_{\text{max}}$ secluded charges denoted by the fields themselves, which corresponds to a solution to eq.~\eqref{eq:Dcoond}, is
\begin{align}
\label{eq:formalsltn}
 \boldsymbol{D} = [\psi_1,\psi_2,\cdots \psi_{N'},\underbrace{\nu,\cdots,\nu}_{\alpha_{\text{max}} = 2\ \text{or}\ 3}]\,.
\end{align}
Without loss of generality, the charges are assumed to be (non-opposite sign) integers, including the dark Higgs charge, $S=\nu/\delta$.  The Yukawa Lagrangian
\begin{align}
    \mathcal{L}_Y = \sum_{ab=1}^{N'} h_{ab}\psi_a\psi_b S^{(*)} + \text{h.c}\,,
\end{align}
must generate a fully massive fermion sector after SSB\footnote{
Note the optional use of the conjugate of $S$ for each Yukawa term, according to the specific charges of the pair $(\psi_a,\psi_b)$.}.
%
A significant advantage of dark sectors protected by gauge symmetries 
is the emergence of a remnant symmetry whenever $|S|>1$. 
In fact, after the SSB, a $Z_{|S|}$ remnant symmetry appears, which ensures the stability 
of the lightest DM candidate~\cite{Batell:2010bp}, 
without needing any extra assumptions.
Given that the minimal number of chiral fermions in the solution ~\eqref{eq:Dcoond} is five~\cite{Costa:2019zzy}, we generally expect a complicated massive fermion sector with multi-component DM candidates. Under these conditions, in ~\cite{Restrepo:2021kpq}, we found one thousand solutions to eq.~\eqref{eq:formalsltn} with 6 to 12 chiral fields and charges of up to 20 in absolute value. Each solution features at least two repeated integers that are assigned as the charges of the right-handed neutrinos. All other chiral fermions in the solution acquire masses after spontaneous symmetry breaking (SSB) through interactions with the dark Higgs. This results in a dark sector with multi-component DM candidates, including at least two-component DM candidates in simpler scenarios~\cite{Restrepo:2021kpq}.

To ensure the two-mediator scenario, we assume that the effective
Dirac neutrino operator is realized only at the radiative level,
mediated by a decoupled scalar inert sector that does not develop
vacuum expectation values. Here, we focus on a two-component DM
model. To maintain a simple inert scalar structure where the neutrino
Dirac masses are generated at one-loop~\cite{Wong:2020obo,
  Bernal:2021ezl} we look for solutions with at least two 
repeated Dirac pairs.  The simplest solution in this case corresponds to
the one with 9 chiral fermions
\begin{align}
    \label{eq:D}
    \boldsymbol{D}=[(1,-10),(1,-10),(-4,-5),9,9,9],
\end{align}
%
and a dark Higgs with a charge of $9$. Notice that one of the
right-handed neutrinos of charge $9$ remains massless. In this
scenario, the first DM candidate is the lightest state of two
Dirac fermions associated with the Dirac pairs $(1,-10)$ and the other DM
candidate is the Dirac fermion associated with the pair
$(-4,-5)$~\cite{Wong:2020obo, Bernal:2021ezl}.

The phenomenological analysis presented in this paper is valid for all
the thousand models found in the previous
work~\cite{Restrepo:2021kpq}, provided that the extra chiral fermions
and inert scalars remain sufficiently decoupled from our two-mediator
and two-component DM scenario.  This applies from the simpler
case with only six chiral fermions: $[5,5,(1,-6),(-2,-3)]$, and a dark
Higgs with charge $5$, with two-component DM candidates but
with a more complicated inert scalar sector~\cite{Bernal:2021ezl}. An
interesting case has 12 chiral fermions and a minimal inert scalar
sector: $[-9, -9, -9, (3), (3), (3),(1, 5),(1, 5), (-7, 13)]$, with a
dark Higgs of charge $6$. In that case, all three right-handed
neutrinos of charge $-9$ generate masses and mixings of three Dirac
neutrinos through a Dirac scotogenic mechanism with Majorana
mediators~\cite{Calle:2019mxn} of charge $3$, which are part of the
three-component and multiflavor DM candidates in the model.

\section{Description of the model}
\label{sec:model}

\begin{table}[ht]
  \centering
  \begin{tabular}{l|c|c|c|c}\hline
    Field                     & Generations & $SU(2)_L$     & $\operatorname{U}(1)_Y$ & $\operatorname{U}(1)_D$  \\ \hline
    $(\nu_{R\alpha})^{\dagger}$ & 3           & $\mathbf{1}$  & $0$                     & $-9$                     \\
    $\chi_{L}$                  & 1           & $\mathbf{1}$  & $0$                     & $4$                      \\ 
    $(\chi_{R})^{\dagger}$      & 1           & $\mathbf{1}$  & $0$                     & $5$                      \\
    $\psi_{Li}$                  & 2           & $\mathbf{1}$  & $0$                     & $10$                     \\ 
    $(\psi_{Ri})^{\dagger}$      & 2           & $\mathbf{1}$  & $0$                     & $-1$                     \\ \hline  
    $S$                         & 1           & $\mathbf{1}$  & $0$                     & $-9$                     \\ \hline 
    $\eta$                      & 1           & $\mathbf{2}$  & $-1/2$                  & $-1$                     \\        
    $\Phi$                      & 1           & $\mathbf{1}$  & $0$                     & $-1$                     \\ \hline 
  \end{tabular}
  \caption{Fermion and scalar content with its quantum numbers.}
  \label{tab:pickedsltn}
\end{table}

Tab.~\ref{tab:pickedsltn} shows new fields added to our model. The first five fields in the table correspond to an anomaly-free set of chiral fermions, singlets under the symmetry group of SM, including the right-handed neutrinos, $\nu_{R \alpha}$. Regarding the scalar fields that acquire VEV in the model, we have $H$, the SM Higgs doublet;  $S$ is an SM singlet scalar that spontaneously breaks the extra Abelian gauge symmetry. The last two fields in the table compose the inert scalar sector required for the scotogenic realization of chiral models, where $\eta$ is a $SU(2)_L$ scalar doublet, $\Phi$ is an SM-singlet scalar. This table also shows the charges under the $U(1)_D$ symmetry. Notice that we have a dark photon corresponding to the dark symmetry. 
This model is a benchmark scenario for chiral DM with scotogenic Dirac neutrino masses with two-mediator and two-component fermionic DM where one Dirac fermion has just one generation and the second DM candidate is the lightest mass eigenstate of mixing with two flavors.
 
The most general Lagrangian, invariant under the SM group and the new $U(1)_D$ symmetry is given by (in two-component notation):
\begin{equation}
\label{eq:Lagrangian}
-\mathcal{L}\supset{y}_{c}\chi_{R}^{\dagger}\chi_{L}S + (y_{x})^{ij}\psi_{Ri}^{\dagger}\psi_{Lj}S +(y_{nR})^{\alpha i}\nu_{R\alpha}^{\dagger}\psi_{Li} \Phi +(y_{nL})^{i\alpha}\psi_{Ri}^{\dagger} L_{\alpha}\,\cdot\, \tilde{\eta} + \text{h.c.}\,,
\end{equation}
where $(\cdot)$ is the $SU(2)_L$ dot product, $\tilde{\eta}=(\eta^+,-\eta^0)^T$, $\alpha,=1,2,3$, $i=1,2.$ and the Yukawa couplings $y$ are assumed real parameters. Also, the Lagrangian contains the scalar potential:
\begin{align}
\label{eq:scalarpotential}
V(H,\eta,S,\Phi) & = -\mu^2\tilde{H}\cdot{H}+m^2_{\eta}\tilde{\eta}\cdot{\eta}+m^2_{\Phi}|\Phi|^2-\mu_S^{2}|S|^2-[\mu_c\tilde{\eta}\cdot{H}\Phi + \text{h.c.}]\nonumber\\
&+\frac{1}{2}\lambda_1(\tilde{H}\cdot{H})^2+\frac{1}{2}\lambda_2(\tilde{\eta}\cdot{\eta})^2+\lambda_3\tilde{H}\cdot{H}\tilde{\eta}\cdot{\eta}+\lambda_4\tilde{H}\cdot{\eta}\tilde{\eta}\cdot{H}+\frac{1}{2}\lambda_5|S|^4\nonumber\\
&+\lambda_6\tilde{H}\cdot{H}|S|^2+\lambda_7|S|^2\tilde{\eta}\cdot{\eta}+\frac{1}{2}\lambda_8|\Phi|^4
+\lambda_9|\Phi|^2\tilde{H}\cdot{H}\,\nonumber\\
&+\lambda_{10}|\Phi|^2|S|^2 +\lambda_{11}|\Phi|^2\tilde{\eta}\cdot{\eta}\,,
\end{align}
where $\mu_k, m_j, \lambda_i$ are real parameters and $\tilde{H}=(0,-\frac{1}{\sqrt{2}}(v+h))^T$.

\subsection{Symmetry breaking and the fermion-scalar spectrum}

\label{subsec:Symmetry_breaking_and_the_fermionscalar_spectrum}

In this model, the VEV $\langle S\rangle=v_s/\sqrt{2}$ of the new scalar field, $S=(S^0+v_s)/\sqrt{2}$, spontaneously breaks the new $U(1)_D$ Abelian gauge symmetry and yields Dirac mass terms for the new dark fermions of the model $\chi_R\,, \chi_L\,, \psi_{R i}\,, \psi_{L j}$. In addition, the Higgs field develops its VEV, $\langle H\rangle=v/\sqrt{2}$, with $v=246.2$ GeV. Consequently, according to the scalar potential~\eqref{eq:scalarpotential}, the Higgs $H$ and the scalar $S$ mix to form two mass eigenstates. In the basis $(h^0,S^0)$, the mass matrix reads
\begin{align}
    m_h^2 =& \left(
\begin{array}{cc}
 -\mu ^2+\frac{1}{2} \lambda _6 v_s^2-\frac{3 \lambda _1 v^2}{2} & \lambda _6 v v_s \\
 \lambda _6 v v_s & -\mu _s^2+\frac{3}{2} \lambda _5 v_s^2+\frac{\lambda _6 v^2}{2} \\
\end{array}
\right)\,,
\end{align}
which is diagonalized by a unitary transformation $Z_H m_h^2Z_H^T = m_{h,\text{diag}}^{2}$, such that:
\begin{equation}
\begin{pmatrix}
h_0 \\ S^0
\end{pmatrix} =
Z_H
\begin{pmatrix}
h_1 \\ h_2
\end{pmatrix}
 =
\begin{pmatrix}
\cos\theta & \sin\theta \\
-\sin\theta & \cos\theta 
\end{pmatrix}
\begin{pmatrix}
h_1 \\ h_2
\end{pmatrix}\,, \label{eq:Higgs-mixing}
\end{equation}
as described in Appendix~\ref{sec:diagonalization-higgs}.

Similarly, after the electroweak and $U(1)_D$ symmetry breaking, the CP-even part $\eta^0$ of the $\eta$ scalar doublet and $\Phi$ mix to form two mass eigenstates. In the basis $(\eta^0,\Phi)$, the Lagrangian reads
\begin{align}
\label{eq:scalarDMmatrix}
\mathcal{L}=\Xi^Tm^2\Xi=\begin{pmatrix}
\eta^0 & \Phi    
\end{pmatrix}\left(
\begin{array}{cc}
 m_{\eta }^2+\frac{1}{2} \lambda _7 v_s^2+\frac{1}{2} \lambda _3 v^2+\frac{1}{2} \lambda _4 v^2 & -\frac{1}{2} v \mu _c \\
 -\frac{1}{2} v \mu _c                                                                      & m_{\Phi }^2+\frac{1}{2} \lambda _{10}
v_s^2+\frac{1}{2} \lambda _9 v^2 \\
\end{array}
\right)\begin{pmatrix}
\eta^0 \\
\Phi
\end{pmatrix}\,,
\end{align}
where the mass matrix ${m}^2_{\Xi}$ can be diagonalized via an orthogonal transformation
\begin{equation}
U_{\Xi}m^2_{\Xi}U^T_{\Xi}=(m^2_{\Xi})^{\text{diag}}=\text{diag}(m^2_1,m^2_2)\,,    
\end{equation}
with
\begin{equation}
U_{\Xi}=\begin{pmatrix}
\cos\theta_{\Xi} & \sin\theta_{\Xi} \\
-\sin\theta_{\Xi} & \cos\theta_{\Xi} 
\end{pmatrix}\,.    
\end{equation}

\begin{figure}[h]
\centering
\includegraphics[scale=0.45]{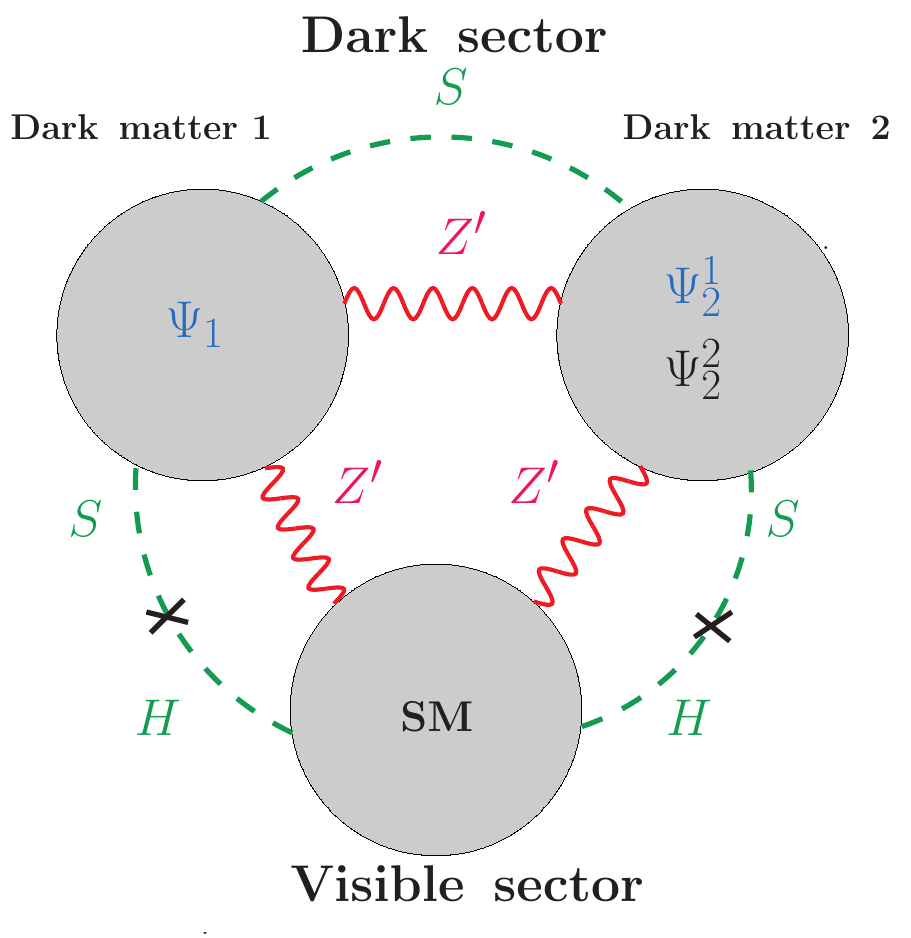}
\caption{A schematic picture that shows our model setup with two secluded DM particles. $\Psi_1$ and $\Psi_2^1$ are the DM candidates. The dark photon $Z^{\prime}$ and the dark Higgs $S$ mediate interactions between the two DM particles. In the case of a non-zero mixing angle between the scalars that acquire VEV or kinetic mixing, the dark sector also interacts with the visible sector.}
\label{fig:dark-conection}
\end{figure}

Regarding the fermion spectrum, in this model, we have two DM particles (DM1 and DM2) that are connected by the dark photon $Z^{\prime}$ and the Higgs portal $S$ as is shown in Fig.~\ref{fig:dark-conection}.  
According to the Lagrangian~\ref{eq:Lagrangian} and after electroweak and $U(1)_D$ symmetry breaking, we get a Dirac fermion DM candidate, $\Psi_1=(\chi_L,\chi_R)^T$, with mass $m_{\Psi_1}=y_c \dfrac{v_s}{\sqrt{2}}$. 
A second DM particle is the lightest eigenstate of two Dirac fermions $(\Psi_2^1, \Psi_2^2)$ that result from the mixing of the four chiral fermions $(\psi_{R}^i)^{\dagger}$ and $\psi_L^j$ $(i,j=1,2)$. Therefore, the Lagrangian in the mass eigenstate basis includes:
\begin{align}
\label{eq:mass-terms}
\mathcal{L}\supset m_{\Psi_1}\overline{\Psi}_1\Psi_1 + \sum_{i=1}^2 m_{\Psi_2^i}\overline{\Psi}_2^i\Psi_2^i\,,  
\end{align}
where $\Psi_1$ and $\Psi_2^1$ are the two DM particles in the model (we assume that $\Psi_2^2$ is heavier than $\Psi_2^1$). Notice that the fermions $\Psi_2^i$ correspond to the eigenstates that are obtained after the diagonalization of the mass matrix $m_{\psi}=(Y_x)^{ij} \dfrac{v_s}{\sqrt{2}}$ with a biunitary transformation $(Z_R, Z_L)$ such that $(m_{\Psi_2}
)^{\text{diag}}=\text{diag}(m_{\Psi_2^1},m_{\Psi_2^2})=Z_Lm_{\psi}Z_R^{\dagger}$ as described in Appendix~\ref{sec:Dirac-neutrino-masses}.
\begin{figure}[h]
\centering
\includegraphics[scale=0.25]{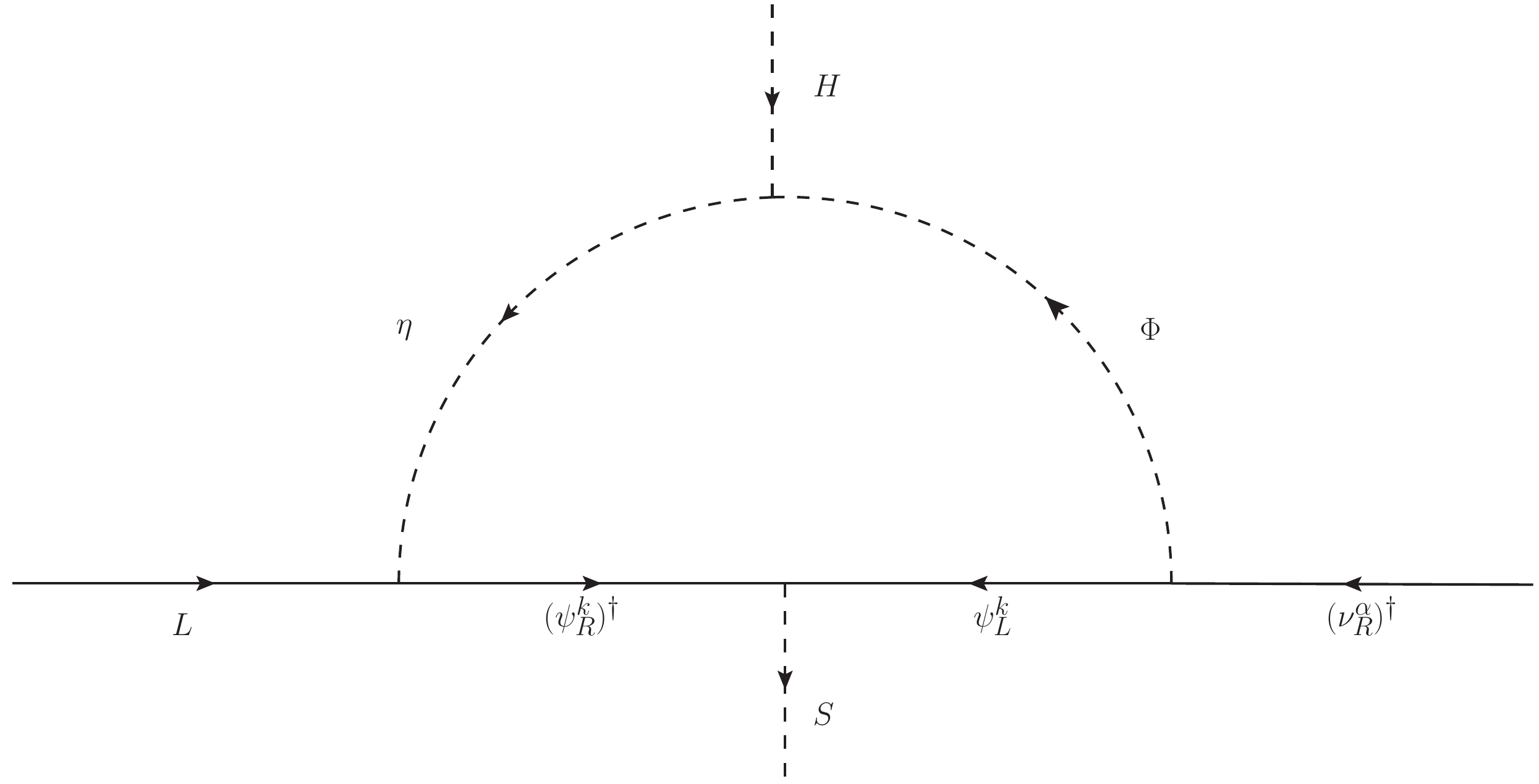}
 \caption{Diagram for neutrino masses at one-loop.}
\label{fig:DiracScotogenic}
\end{figure}
Finally, we have neutrino masses by realizing the effective Dirac neutrino masses operator shown in Eq.~\eqref{eq:neutrino-operator} (with $\delta=1$). The specific realization is shown in Fig.~\ref{fig:DiracScotogenic} and, in Appendix~\ref{sec:Dirac-neutrino-masses}, we obtain all the Yukawa couplings in this Feynman diagram that are compatible with neutrino oscillation data to $3\sigma$~\cite{deSalas:2017kay}.

\subsection{Kinetic mixing and the electroweak gauge bosons}

\label{subsec:Kinetic_mixing_and_the_electroweak_gauge_bosons}

This model has a new gauge boson $B'_{\mu}$ associated with the $U(1)_D$ gauge symmetry. It interacts with the SM sector via kinetic mixing $\epsilon$ parameter. The Lagrangian for $B'_{\mu}$ reads~\cite{Cline:2014dwa}
\begin{align}
\label{eq:La_DM}
\mathcal{L}_{B'}=&-\frac{1}{4}B'_{\mu\nu}B^{'\mu\nu}-\frac{\epsilon}{2}B'_{\mu\nu}B^{\mu\nu}
+\frac{i}{2}g_D\overline{\Psi}_1\gamma^{\mu}(a_1+b_1\gamma^5){B}'_\mu\Psi_1\nonumber\\
&+\frac{i}{2}g_{\operatorname{D}}\sum_{k=1}^2 \overline{\Psi}_2^k\gamma^{\mu}(a_2+b_2\gamma^5){B}'_\mu\Psi_2^k+(D^X_{\mu}X)^{\dagger}D^{X\mu}X+ 9i\sum_{\alpha=1}^2\overline{\nu}_{\alpha}\gamma^{\mu}(1+\gamma^5){B}'_\mu\nu_{\alpha}\,,
\end{align}
where, $a_k=(q_{\psi^k_L}-q_{\psi^k_R})$, $b_k=(q_{\psi^k_L}+q_{\psi^k_R})$,  $B'_{\mu\nu}=(\partial_{\mu}B'_{\nu}-\partial_{\nu}B'_{\mu})$,  $B_{\mu\nu}=(\partial_{\mu}B_{\nu}-\partial_{\nu}B_{\mu})$ are the strength tensors for $B'_{\mu}$ and $B_{\mu}$ respectively, and
\begin{equation}
\mathcal{D}^X_{\mu}=\partial_\mu-iq_Xg_DB'_{\mu}\,,
\end{equation}
for $X=S,\Phi,\eta$ and $q_X$ are the charges under $U(1)_D$ gauge symmetry, $g_D$ is the new $U(1)_D$ gauge coupling and the $\nu_{\alpha}$ are Dirac neutrinos.
In the basis $V_{\mu}=(B_{\mu}, W^3_{\mu}, B'_{\mu})^{T}$,  the mass matrix for the neutral gauge bosons reads~\cite{Bauer:2018onh}
\begin{equation}
\mathcal{L}_M = \frac{1}{2} V^{T\mu}M^2_GV_{\mu}\,,
\end{equation}
where
\begin{align}
\label{eq:gauge_boson_mass_matrix}
M^2_G = \frac{1}{4}v^2 
\begin{pmatrix}
 g_1^2 & -g_1 g_2  & - g_1^2 \epsilon  \\
 - g_1 g_2 & g_2^2 &  g_1 g_2\epsilon  \\
 - g_1^2  \epsilon  &  g_1 g_2 \epsilon  & 324g_D^2\frac{v^2_S}{v^2}+g_1^2\epsilon ^2 \\
\end{pmatrix}\,.
\end{align}
After the diagonalization of this matrix, we have three eigenstates. Those are the massless $\gamma$ photon, the SM $Z$ gauge boson with a mass $m_Z\approx 91.1$ GeV, and the new dark photon $Z'$ with a mass $m_{Z'}\approx 9g_D v_s/(1+\epsilon^2)$.

\section{Dark matter freeze-out} 
\label{sec:DM}
\begin{figure}[h]
\centering
\includegraphics[scale=0.4]{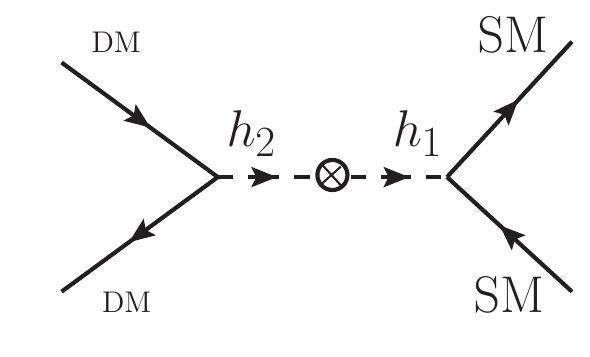}
\includegraphics[scale=0.4]{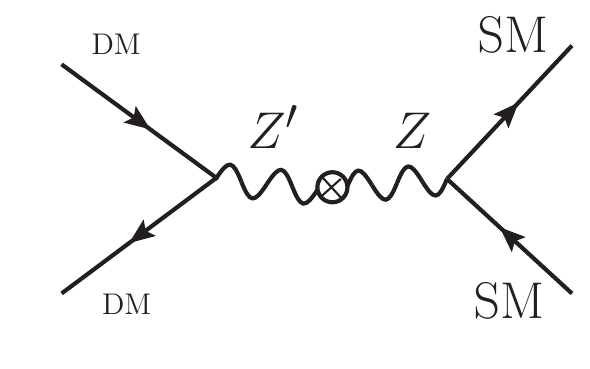}
\includegraphics[scale=0.4]{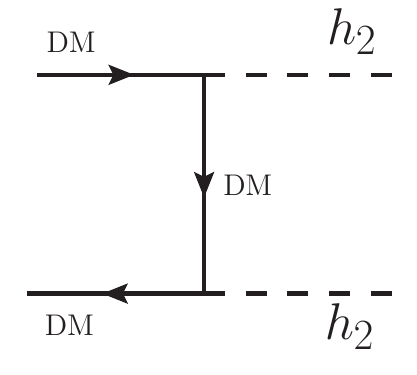}
\\
\includegraphics[scale=0.4]{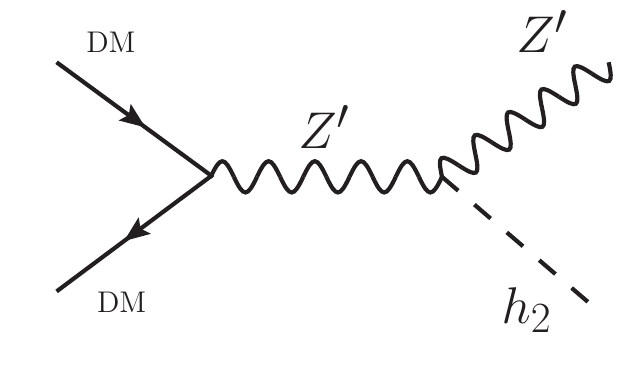}
\includegraphics[scale=0.4]{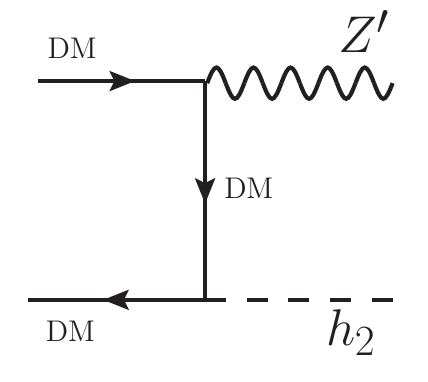}
\includegraphics[scale=0.4]{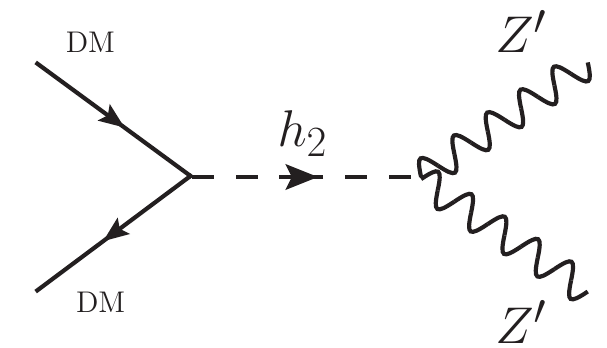}
\includegraphics[scale=0.4]{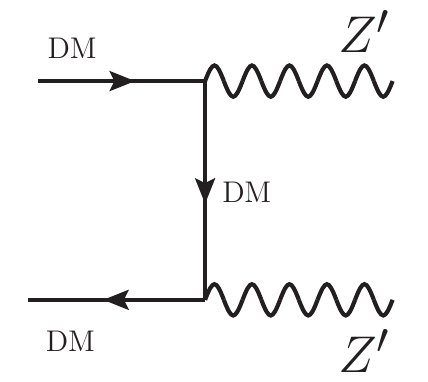}
\caption{$\text{DM}(\Psi_1, \Psi_2^1)$ annihilation channels (we do not show the corresponding u-channels.). The DM annihilation into SM model particles via mass mixing of the scalars $h_i$ or kinetic mixing is extremely suppressed.}
\label{fig:DMDMSMSM}
\end{figure}
\begin{figure}[h]
\centering
\includegraphics[scale=0.5]{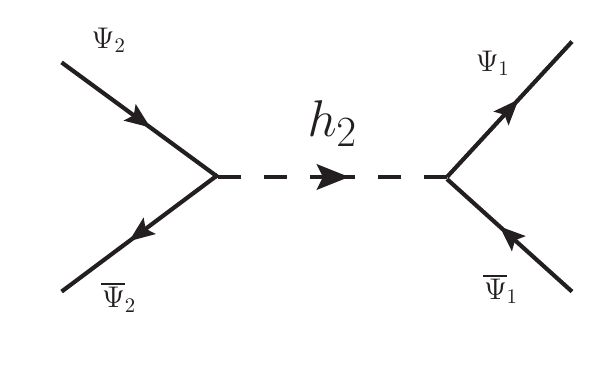}
\includegraphics[scale=0.5]{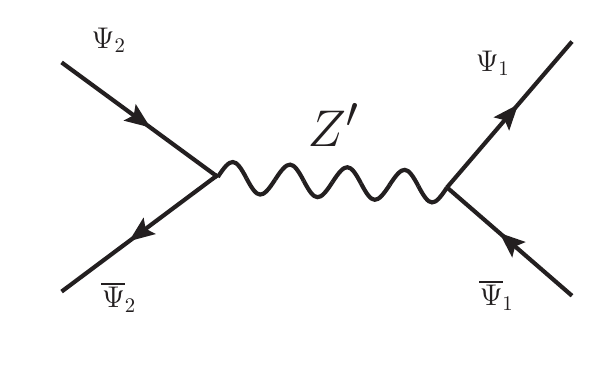}
\caption{DM conversion channels $\Psi_1 \leftrightarrow \Psi_2^1$ mediated by the dark Higgs $h_2$ or the dark photon $Z^{\prime}$.}
\label{fig:DMconversion}
\end{figure}

In this model, the $U(1)_D$ dark symmetry determines the types of processes that render the relic abundance of DM. In Fig.~\ref{fig:DMDMSMSM}, we show the direct annihilation into SM particles, Dirac neutrinos, the dark Higgs $h_2$, and the dark photon $Z^{\prime}$. Also, we show the DM conversion processes in Fig.~\ref{fig:DMconversion}.

The complete set of $2\to 2$ processes that contribute to the relic density of DM can be classified by four digits~\cite{Belanger:2014vza}. For example, $0$ is used for SM particles including the $Z'$ and the $h_2$ fields, $1$ for the $\Psi_1$ and $\overline\Psi_1$, $2$ for $\Psi_2$ and $\overline\Psi_2$. 
For instance, the processes that are classified in the type $1100$ have two SM particles in the final state and a pair $\Psi_1$, $\overline\Psi_1$ in the initial state. 
Notice that the $U(1)_D$ dark symmetry forbids the semi-annihilation process as $1120$ (only one SM particle) and the conversion process as $1112$. Table~\ref{tab:type-processes} shows all the processes that contribute to the relic density of DM with their respective classification.
\begin{table}[h]
    \centering
    \begin{tabular}{c|c}
         Processes &Type  \\
         \hline
         $\overline\Psi_1$ $\Psi_1$ \hspace{0.2 cm}$\to$\hspace{0.2 cm} SM SM & 1100 \\
         $\overline\Psi_2^1$ $\Psi_2^1$ \hspace{0.2 cm}$\to$\hspace{0.2 cm} SM SM & 2200\\
         $\overline\Psi_1$ $\Psi_1$ \hspace{0.2 cm}$\to$\hspace{0.2 cm} $\overline\Psi_2^1$ $\Psi_2^1$ & 1122 \\
            $\overline\Psi_2^1$ $\Psi_2^1$ \hspace{0.2 cm}$\to$\hspace{0.2 cm} $\overline\Psi_1$ $\Psi_1$ & 2211
    \end{tabular}
    \caption{The $2\to 2$ processes allowed in this model that can modify the relic density of DM particles.}
    \label{tab:type-processes}
\end{table}

The relic abundance of DM $\Omega_i$ $(i=1,2)$ for ($\Psi_1, \Psi_2^1$) is obtained by solving the Boltzmann equations:
\begin{align}
\label{eq:boltzmann-equations}
\frac{\operatorname{d}n_1}{\operatorname{d}t}= & -\sigma^{1100}_v(n^2_1-\overline{n}^2_{1})-\sigma^{1122}_v\left(n^2_1-n^2_2\frac{\overline{n}^2_{1}}{\overline{n}_2}\right)-3Hn_1\,,\nonumber\\
\frac{\operatorname{d}n_2}{\operatorname{d}t}= & -\sigma^{2200}_v(n^2_2-\overline{n}^2_{2})   -\sigma^{2211}_v\left(n^2_2-n^2_1\frac{\overline{n}^2_{2}}{\overline{n}_1}\right)-3Hn_2\,,
\end{align}
where $n_i\, (i=1,2)$ is the number density for the DM particle, $\overline{n}_i$ are their respective equilibrium values and $\sigma_v^{abcd}$ is the thermally averaged cross section, that satisfies the relation $\overline{n}_a\overline{n}_b\sigma_v^{abcd}=\overline{n}_c\overline{n}_d\sigma_v^{cdab}$.

In the next section, we compute the analytical expression for the relic density in a simplified regime. However, in this work, we used \texttt{micrOMEGAs 6.0.3}~\cite{Alguero:2023zol}  to compute $\Omega_i$ because it considers all the processes involved in the thermal evolution of the coupled Boltzmann eqs.~\eqref{eq:boltzmann-equations}.
We define the following parameters for the DM particles that allow us to study the relevant processes that play a role in the evolution of the Boltzmann equations and consequently affect the relic abundance of DM:
\begin{align}
\label{eq:zeta-parameters}
    \zeta^i_{anni}(T)=&\dfrac{\sigma_v^{ii00}(T)}{\sigma_v^{ii00}(T)+\sigma_v^{iijj}(T)}\\
    \zeta^i_{conv}(T)=&\dfrac{\sigma_v^{iijj}(T)}{\sigma_v^{ii00}(T)+\sigma_v^{iijj}(T)}\,,
\end{align}
$i\neq j=1,2$, where those parameters are evaluated at the typical freeze-out temperature: $T\approx m_{\chi_1}/25$.  Notice that by construction, 
 $\zeta^i_{anni}+\zeta^i_{conv}=1$, and for example: $\zeta^1_{anni}\approx 1$ means that the relic density of $\Psi_1$ particle is dominated by the annihilation process $1100$ and the conversion $1122$ ($2211$) contributions are negligible. In practice, to compute those parameters we used the specific function \texttt{vsabcdF}$(T)$ incorporated into the last version of \texttt{micrOMEGAs 6.0.3}~\cite{Alguero:2023zol}, which allows us to evaluate the cross sections $\sigma_v^{iijj}, \sigma_v^{ii00}$ at some specific temperature.

Finally, we impose the constraint:
\begin{equation}
\label{eq:Omega}
\Omega h^2=\Omega_1h^2+\Omega_2h^2 = 0.1200\pm 0.0012 \,,   
\end{equation}
where $\Omega h^2$ is the observed value by the PLANCK experiment~\cite{Planck:2018vyg}. Notice that, experimental signals of DM in any detector are affected by the fraction of the local abundance of the DM candidate. This fraction is defined as:
\begin{align}
\label{eq:xi}
    \xi_i = \dfrac{\Omega_i}{\Omega}\,, (i=1,2)\,,
\end{align}
such that $\xi_1+\xi_2=1$. This feature distinguishes between a case with one single-component DM candidate and a multicomponent scenario. 

Since the dark sector is in thermal equilibrium, the abundance of DM is determined by the thermal freeze-out into lighter hidden particles $h_2$ and $Z^{\prime}$ as shown in Fig.~\ref{fig:DMDMSMSM}.
Notice that $h_2$ is a new Higgs scalar that could be detected in high energy experiments. Also, notice that the dark photon $Z^{\prime}$ is unstable because it decays into Dirac neutrinos with a decay width
\begin{align}
\label{eq:Zp-dacay}    
    \Gamma_{Z'\to\overline{\nu}_i\nu_i} = 3\,M_{Z'}\dfrac{(g_1^2\epsilon^2+4\,q_{\nu_R}^2\,g_D^2\,)}{64\pi} \,,
\end{align}
where $q_{\nu_R}=9$ is the charge of the right-handed neutrino (see Tab.~\ref{tab:pickedsltn}). 
Experimentally, the dark photon $Z^{\prime}$ needs to decay in less than one second, $1/\Gamma \lesssim 1$ sec, and is absent at the time of Big Bang Nucleosynthesis (BBN)~\cite{Pospelov:2007mp}. 
Since left-handed neutrinos belong to the visible sector (VS) and right-handed neutrinos are in the dark sector (DS) in the early Universe, the decay of $Z^{\prime}$ does not reheat SM plasma. 
The visible and the dark sector are in thermal equilibrium unless the rate of interaction for the processes that communicate the two sectors $\Gamma$ is much smaller than the Hubble rate $H$, $\Gamma\ll{H}$. That occurs when the kinetic mixing is tiny $\epsilon\sim{10}^{-6}$, the small mixing angle between the SM Higgs and the dark Higgs is also tiny $\theta<10^{-3}$, and the scalars that allow the generation of neutrino masses are very heavy $m_{\Xi}>1\, \text{TeV}$ as shown in Ref.~\cite{Wong:2020obo}.

%
Notice that our model could render the relic abundance of DM compatible with BBN even without the kinetic mixing parameter~\cite{Hannestad:1995rs, Kawasaki:2000en, Coy:2024itg}. Also, cosmological constraint due to the contribution of right-handed neutrinos to relativistic degrees of freedom in the early Universe is satisfied as shown in Appendix~\ref{appendix:Delta_N_Effective}. Thus, a fully consistent scenario, even without kinetic mixing, is compatible with all theoretical, cosmological, and phenomenological constraints. Finally, we emphasize that at low temperatures, left-handed and right-handed neutrinos mix to form Dirac neutrinos in a process known as ``left-right equilibration"~\cite{Chen:2015dka}. 

\subsection{One-component WIMP Limit}

\label{subsec:Onecomponent_WIMP_Limit}

As a benchmark scenario, suppose that $m_{Z^{\prime}}<m_{\Psi_1}\ll m_{\Psi_2^1}$, the kinetic mixing and the mixing angles $\theta$ in the scalar sector are zero, and $ m_{h_2}>2m_{\Psi_1} \gg 125$ GeV. In this scenario, Boltzmann eqs.~\eqref{eq:boltzmann-equations} disentangle, and the relic density associated with the DM particle that has only one generation domains the total relic density ($\zeta_{conv}^1\to 0$). 
In this scenario, the DM abundance is rendered by secluded thermal freeze-out, through $\overline\Psi_1\Psi_1\to Z'Z'$ processes shown in Fig.~\ref{fig:DMDMSMSM} (notice that $\overline\Psi_1\Psi_1\to Z'h_2$ is forbidden by kinematics). 

In this limit, the thermal evolution of $\Psi_1$ follows the freeze-out mechanism in the dark sector. As the Universe adiabatically cooled down, the Universe's expansion overtook the DM annihilation rate $\Gamma \ll H$, and the relic density of DM was frozen out. The Boltzmann eqs.~\eqref{eq:boltzmann-equations}, yields~\cite{Kolb:1990vq, Srednicki:1988ce}
\begin{equation}
\label{eq:RelicAbundance}
\Omega_1 h^{2} \approx \frac{2.08 \times 10^{9}\, x_{f}\,\,\text{GeV}^{-1}}{M_{\mathrm{Pl}} \sqrt{g_{*}(T_f)}\left(a+3 b / x_{f}\right)}\,,
\end{equation}
where the parameters $a$ and $b$ comes from the thermally-averaged annihilation cross-section $\langle \sigma v\rangle = (a\,+\,b v^2\,+\,\mathcal{O}(v^4))$, $g_*(T)$ is the effective number of degrees of freedom at the temperature $T$, $x_f= m_{\Psi_{1}}/T_{\rm freeze-out}$ and where $h$ is today's Hubble parameter in units of $100~{\rm km/s/Mpc}$.
Also, in this limit, the thermally averaged annihilation cross section is given by (see Eq.\eqref{eq:a-Model}):
\begin{equation}
\label{eq:sigmav}
\langle\sigma{v}\rangle\approx 
\frac{g_D^4 \left(1-r^2\right)^{3/2} \left(1519\, r^2+162\right)}{16\, \pi\,  m_{ \Psi_1}^2 r^2 \left(r^2-2\right)^2}
+b\,v^2\,,
\end{equation}
with $r=m_{Z^{\prime}}/m_{\Psi_1}$ and the p-wave is given by Eq.\eqref{eq:b-Model}.
\footnote{In the appendix~\ref{sec:general-sv} we computed the $\langle\sigma{v}\rangle$ with \texttt{FeynArts 3.11}~\cite{Hahn:2000kx} and \texttt{FeynCal 10.0.0}~\cite{Mertig:1990an, Shtabovenko:2016sxi, Shtabovenko:2020gxv}. We obtained a general expression that matches the vector-like limit studied in Refs.~\cite{Pospelov:2007mp, Ma:2021szi, Ma:2022uhi, delaVega:2023dmw} and the axial-vector case in Refs.\cite{Bell:2016uhg, Alves:2015pea,Alves:2015mua}} 

\begin{figure}[h]
\centering
\includegraphics[scale=0.7]{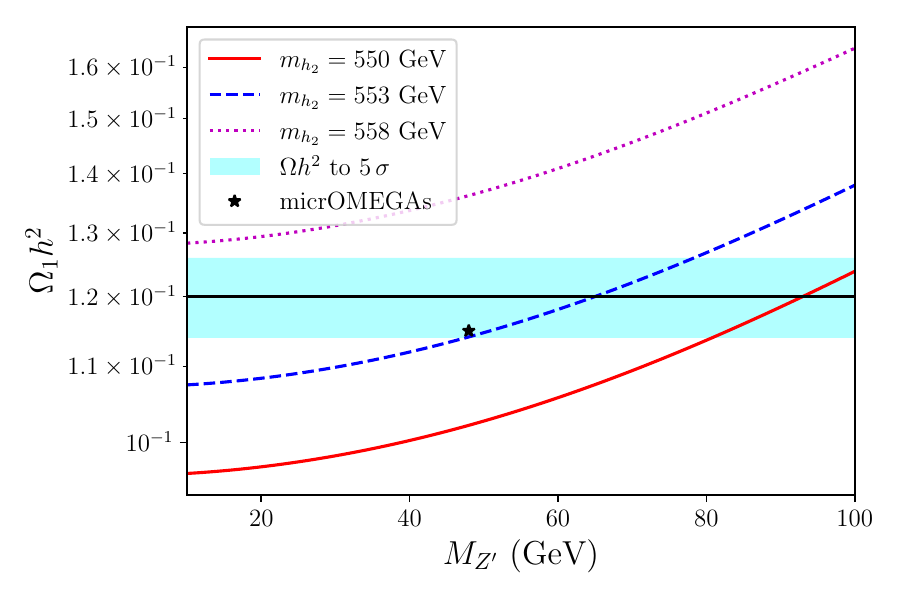}
 \caption{Relic density in the limit of one DM component $\Psi_1$.}
 \label{fig:benchmark-DMDMZpZp}
 \end{figure}

In Fig.~\ref{fig:benchmark-DMDMZpZp}, we show the behavior of the relic abundance $\Omega_1 h^2$ given by the Eq.~\eqref{eq:RelicAbundance} as a function of the dark photon mass $m_{Z'}$. We choose the benchmark scenario with $m_{\Psi_2^1}=1000\, \text{GeV}$, $m_{\Psi_2^2}=2000\, \text{GeV}$, $y_c=0.85$, $v_s=400\, \text{GeV}$,  $1<m_{Z^{\prime}}/\text{GeV}< 100$ and $m_{h_2}=\{550, 553, 558\}\, \text{GeV}$. 
The black star in Fig.~\ref{fig:benchmark-DMDMZpZp} is the benchmark point for $m_{Z^{\prime}}\approx 43$ GeV and $m_{h_2}=553$ GeV, computed with ~\texttt{micrOMEGAs} that shows a good agreement with the limit of one DM component. For this point $\langle \sigma v\rangle\approx 3\times 10^{-26}$ cm$^3/$s. 

\section{Phenomenology of the model}

\label{sec:phenomenology}

\begin{table}[ht]
  \centering
  \begin{tabular}{l|l}\hline
    Parameter & Range\\\hline
    $M_{Z'}/\text{GeV}$ & $1-10^3$ \\
    $g_D$ & $10^{-3}-1$ \\
    $y_c$ & $10^{-3}-1$\\
    $(m_{\Psi_2^j}-m_{\Psi_1})/\text{GeV}$ & $1-5\times 10^3$ \\
    $\theta$ & $10^{-6}-10^{-3}$ \\
    $\theta_{L}, \theta_{R}$ & $10^{-3}-2\pi$ \\
    $\lambda_{k}$ &  $10^{-4}-1$\\
    $m_{h_2}/\text{GeV}$ & $125-5\times 10^3$ \\
    $m_{\eta}^2/\text{GeV}^2$ & $10^6-10^8$\\
    $m_{\Phi}^2/\text{GeV}^2$ & $10^6-10^8$\\
    $\mu_c/\text{GeV}$ & $10^2-2\times 10^3$\\
    $\epsilon$ & $10^{-12}-10^{-2}$\\
    $y_{nL}$ & $10^{-4}-1$ \\
    \hline
  \end{tabular}
  \caption{Scan ranges for the free parameters in this model. $k=\{2,\cdots 11\}$, $k\neq\{5,6\}$, $j={1,2}$.}
  \label{tab:scan}
\end{table}
We implemented this model in \texttt{SARAH}~\cite{Staub:2008uz,Staub:2009bi,Staub:2010jh,Staub:2012pb,Staub:2013tta}, and coupled it with the \texttt{SPheno}~\cite{Porod:2003um,Porod:2011nf} routines to obtain the spectrum of the model. We computed the DM relic density using \texttt{micrOMEGAs 6.0.3}~\cite{Alguero:2023zol} that numerically solves the coupled set of Boltzmann eqs.~\eqref{eq:boltzmann-equations} for the two DM candidates. We selected the points in the parameter space compatible with the current value of the DM relic density in eq.~\eqref{eq:Omega} and the neutrino masses (see Appendix~\ref{sec:Dirac-neutrino-masses}). 
We also considered the constraints due to the branching ratio of Higgs and $Z$ boson decays into invisible as shown in Appendices~\ref{appendix:Higgs_Decays}, \ref{appendix:Z_Decays}. The branching ratios of the decays of Higgs and Z bosons into invisible particles are many orders of magnitude below the current constraint and do not affect our model's parameter space. However, cosmological constraints due to the effective number of relativistic degrees of freedom in the early Universe constrain the value of the kinetic mixing of the model as shown in Appendix~\ref{appendix:Delta_N_Effective}.

We randomly scanned the model's parameter space by varying the parameters as shown in Table~\ref{tab:scan}. 
The parameters in the model were chosen as real. We fixed the SM Higgs mass $m_{h_1}=125$ GeV, $m_{h_2}>m_{h_1}$ and $\lambda_1, \lambda_5, \lambda_6 $ were computed in terms of the mixing angle $\theta$ as described in the appendix~\ref{sec:diagonalization-higgs}.
According to collider constraints, it follows that $|\sin\theta|< 0.3$ for $m_{h_2}>m_{h_1}$~\cite{Falkowski:2015iwa,Arcadi:2019lka,Ferber:2023iso,CMS:2022dwd}.
However, in this work, we considered $10^{-6} <\theta < 10^{-3}$ which suppresses the dark Higgs portal for direct detection of DM. This value for $\theta$ was motivated because the DM is a Dirac particle in this model, and the vector portal is open for direct detection. For Majorana DM, instead the dark Higgs portal is the only open channel for direct detection as shown in Ref.~\cite{Babu:2024zoe}.
Also, the Yukawa couplings $(y_x)^{ij}$ in Eq.~\eqref{eq:Lagrangian} were computed in terms of $m_{\Psi_2^1},m_{\Psi_2^2}$ and the mixing angles $\theta_{L,R}$ as described Appendix~\ref{sec:Dirac-neutrino-masses}. 
Finally, the Yukawa couplings $(y_{nR})^{\alpha i}$ in Lagrangian~\eqref{eq:Lagrangian} were parameterized in terms of the $(y_{nL})^{i\alpha}$ Yukawa couplings, the neutrino masses $m_{\nu \alpha}$ and the Pontecorvo-Maki-Nakagawa-Sakata matrix~\cite{Maki:1962mu} as described in Appendix~\ref{sec:Dirac-neutrino-masses}. In this model, the new Yukawa couplings reproduce the current neutrino oscillation data by construction~\cite{deSalas:2017kay}.

\begin{figure}[h]
 \centering
\includegraphics[scale=0.6]{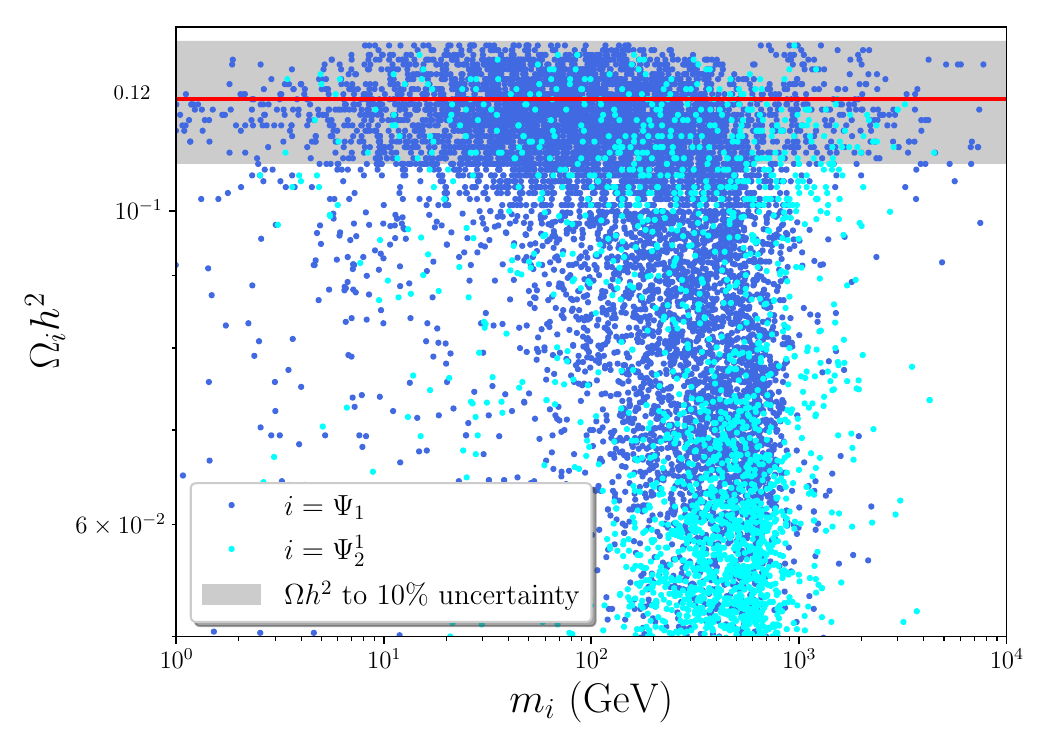}
 \caption{DM relic density for the two fermionic DM candidates ($\Psi_1$, $\Psi_2^1$).}
\label{fig:DM}
\end{figure}

In this paper, we study only the case of fermionic DM. Notice that the lightest mass eigenstate of mixing between $\eta$ and $\Phi$ could be a scalar DM candidate. However, in this setup, we consider a heavy scalar with a mass larger than $1$ TeV as shown in Tab.~\ref{tab:scan}. Fig.~\ref{fig:DM} shows the behavior of the DM relic density for the two fermionic DM candidates. 
Notice that our model can render DM relic abundance for masses that range between $1$ GeV to the TeV scale. Each fermionic DM candidate ($\Psi_1$, $\Psi_2^1$) could render the $100\%$ of the DM abundance, those are the points in the grey horizontal band with $10\%$ of uncertainty and $\xi_i=(\Omega_i/\Omega)\approx 1$. In addition, the interaction between the two candidates for DM gives the value $\Omega h^2\approx 0.12$~\cite{Planck:2018vyg}. They correspond to the models below the grey band with $\xi_i=(\Omega_i/\Omega)\leq 1$, but $\xi_1+\xi_2=1$. Also, we realize that for a DM masses bigger than $m_{Z^{\prime}}$, the process $\overline{\Psi}\Psi\to Z'Z'$ is kinematically favored and domains the scan with a branching near to one (as shown in s, u and t-channels in Fig.~\ref{fig:DMDMSMSM}). 
In contrast, a DM particle lighter than $m_{Z^{\prime}}$ decays directly into SM fermions, preferably into two Dirac neutrinos $\overline{\Psi}\Psi\to \overline{\nu}_i\nu_i$ ($i=1,2,3$).  
 \begin{figure}[h]
 \centering
\includegraphics[scale=0.45]{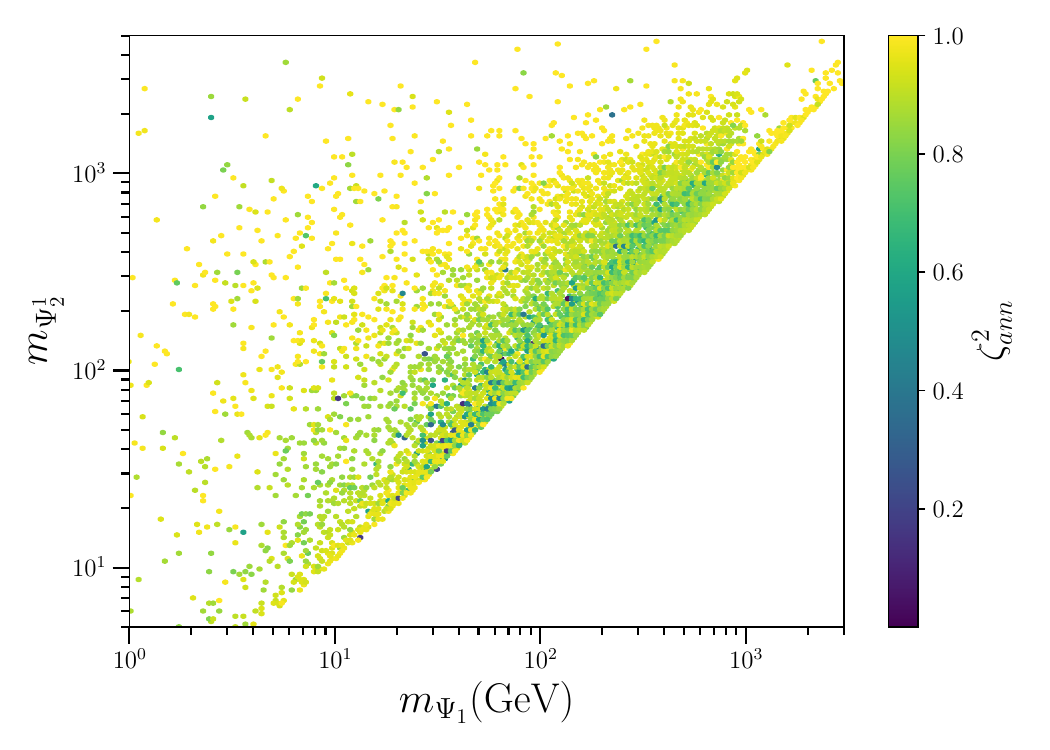}
\includegraphics[scale=0.45]{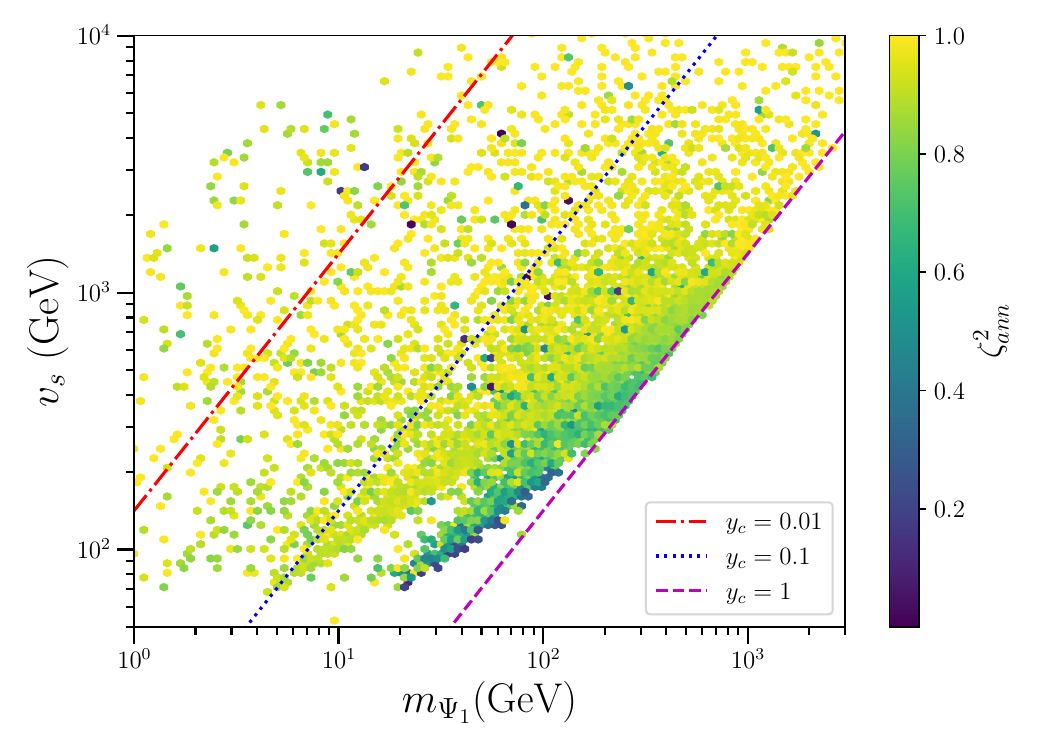}
 \caption{DM conversion impact. The green darker points represent models where the conversion thermal cross-section $\sigma_v^{2211}$ significantly contributes to the relic abundance of DM.}
\label{fig:zeta-ann}
\end{figure}

Also, we study the DM conversion described by the Feynman diagram shown in Fig.~\ref{fig:DMconversion}. In Fig.~\ref{fig:zeta-ann} we show the effect of the type process 2211 in Tab.~\ref{tab:type-processes}. Notice that DM conversion $\overline{\Psi}_2\Psi_2 \to \overline{\Psi}_1\Psi_1$ is kinematically allowed because $m_{\Psi_1}<m_{\Psi_2^1}$ as we described in Tab.~\ref{tab:scan}. 
These plots show that the thermal annihilation cross section $\sigma_v^{2211}$ in Bolztmann eqs.~\ref{eq:boltzmann-equations}, which is included in the parameter $\zeta_{ann}^2(T)$, contributes significantly to the relic abundance of DM for some points of our DM model (green points).

In the left plot of Fig.~\ref{fig:zeta-ann} we show the behavior of the $\zeta^2_{ann}$ parameter in the $|m_{\Psi_1} - m_{\Psi_2^1}|$ plane (see eq.~\ref{eq:zeta-parameters}). The relic abundance of DM is almost always dominated by annihilation processes directly to the SM particles where $\zeta_{ann}^2\approx 1$ (yellow points). However, for the green darker points, the DM conversion process $\overline{\Psi}_2\Psi_2 \to \overline{\Psi}_1\Psi_1$ is relevant and the thermal cross section $\sigma_v^{2211}$ plays an important role in rendering the relic abundance of DM.
Also, in the right plot of Fig.~\ref{fig:zeta-ann} we show the behavior of the $\zeta^2_{ann}$ parameter in the $|m_{\Psi_1} - v_s|$ plane. The Yukawa coupling $y_c$ in the Lagrangian~\ref{eq:Lagrangian} not only determines the mass of the lightest DM particle but also plays an important role in the DM conversion process $\overline{\Psi_2}\Psi_2 \to \overline{\Psi_1}\Psi_1$ that directly impacts the relic abundance $\Omega_1$. This conversion process is activated principally for $y_c\gtrsim 0.1$ and $v_s\lesssim 1$ TeV (dark green points). 

\subsection{Direct and indirect detection of DM}
\label{sec:direct-detection}

\begin{figure}[h]
\centering
\includegraphics[scale=0.6]{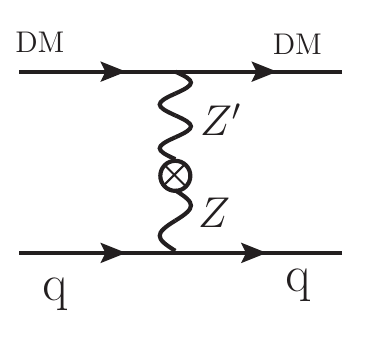}
\includegraphics[scale=0.6]{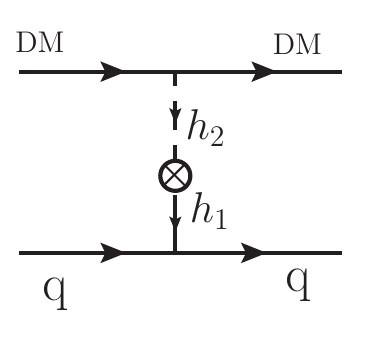}
\caption{Spin independent interactions of DM $(\Psi_1, \Psi_2^1)$ with nuclei: vector (left) and scalar (right) portals.}
\label{fig:SI-diagrams}
\end{figure}
This model allows for direct detection signatures since the DM scatters with nuclei through
the t-channel in the vector and scalar portal shown in Fig.~\ref{fig:SI-diagrams}.
The contribution of the vector portal (left) to the spin independent (SI) cross-section is given by~\cite{Duerr:2014wra, Restrepo:2022cpq}
\begin{align}
\label{eq:SI-vector}
    \sigma_{\Psi_{iN}}^{\text{SI-vector}}=&\frac{\mu^{2}}{4\pi}\frac{g_D^2g^2_1\epsilon^2}{M^4_{Z'}}B_{\Psi_i}^2 \,, 
\end{align}
where $\mu=M_N m_{\Psi_i}/(M_N+m_{\Psi_i})$ is the reduced mass,  $M_N\approx939$ MeV is the nucleon mass (neutron or proton), $B_{\Psi_1}=(q_{\chi_L}-q_{\chi_R})=1$, $B_{\Psi_2}=(q_{\psi_L}-q_{\psi_R})=9$ (see Tab.~\ref{tab:pickedsltn}).
On the other hand, the contribution of the scalar portal (right) to the spin SI cross-section is given by~\cite{Ferber:2023iso,Yaguna:2024jor}:
\begin{align}
\label{eq:SI-scalar}
    \sigma_{\Psi_{iN}}^{\text{SI-scalar}}=&
    \dfrac{\mu^2}{2\pi}\left(\dfrac{y'\cos{\theta}\sin\theta}{v}\right)^2\left(\dfrac{1}{m_{h_2}^2}-\dfrac{1}{m_{h_1}^2}\right)^2 f^2 M_N^2\,, 
\end{align}
with the effective Higgs-nucleon coupling $f\approx 0.3$ and $y'$ is the scalar coupling between the dark Higgs and the DM particle. 
After the scan was done, we noticed that the vector interaction was dominant. 
The scalar portal is subdominant because the Higgs mixing angle $\theta$ was chosen less than $10^{-3}$ in Table~\ref{tab:scan}. 
Note that although kinetic mixing $\epsilon$ plays an important role in the direct detection of DM and opens the door to current and future searches of DM, it does not play a significant role in determining the relic abundance of DM in this secluded framework.

\begin{figure}[h]
\centering
\includegraphics[scale=0.5]{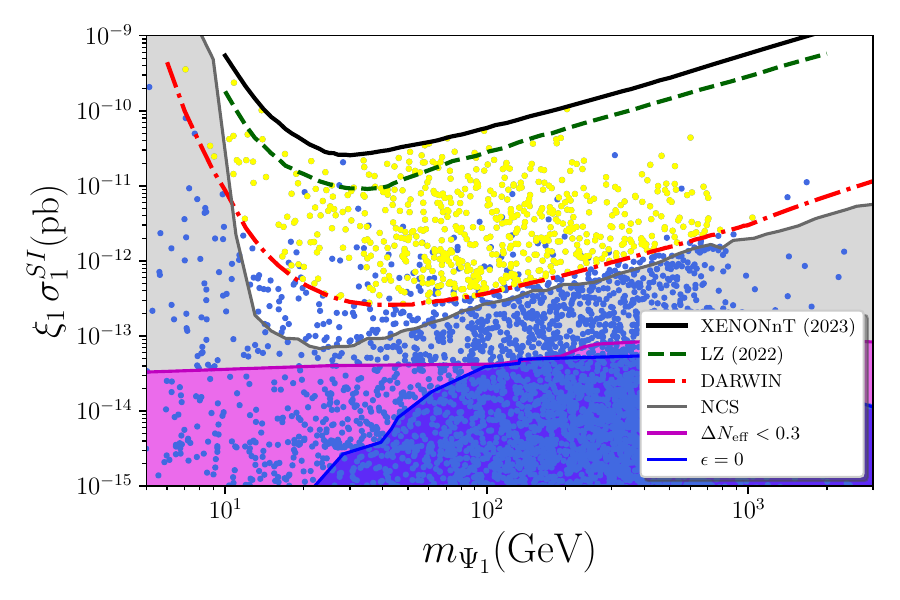}
\includegraphics[scale=0.5]{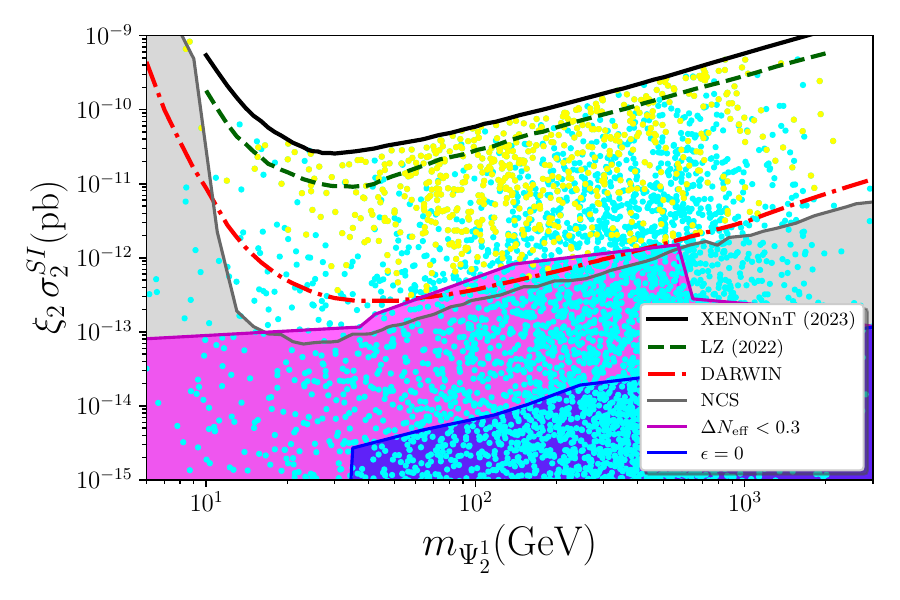}
 \caption{SI cross-section for elastic scattering of DM with nuclei scaled by $\xi_i$ fraction of DM. We also show the limit set by XENONnT collaboration~\cite{XENON:2023cxc}, LZ~\cite{LZ:2022lsv} and the projected WIMP sensitivity from DARWIN~\cite{DARWIN:2016hyl} experiments. Also, we show the Neutrino Coherent Scattering (NCS)~\cite{Cushman:2013zza, Billard:2013qya}, the allowed region by the $\Delta N_{\text{eff}}$ (magenta color) and the region with $\epsilon=0$ (blue color). Yellow points represent the viable models where both DM particles are expected to yield signals in future experiments.}
\label{fig:SIs}
\end{figure}

Fig.~\ref{fig:SIs} shows the SI cross section for elastic scattering of DM with nuclei for each DM component of the model ($\Psi_1$ on the left and $\Psi_2^1$ on the right.). The SI scales with the parameter $\xi_i=\Omega_i/\Omega$ that accounts for the DM density fraction on Earth (see eq.~\eqref{eq:xi}).
After we scan the parameter space of the model for parameters that range as shown in Tab.~\ref{tab:scan}, we filter all the points (models) that fall under the current limit set by XENONnT experiment~\cite{XENON:2023cxc}. We also show the results of the LUX-ZEPLIN (LZ) experiment~\cite{LZ:2022lsv} and the prospects of the DARWIN experiment~\cite{DARWIN:2016hyl} that is one of the final next generation of DM experiments before reaching the threshold for Neutrino Coherent Scattering (NCS)~\cite{Cushman:2013zza, Billard:2013qya}, where the neutrino-nuclei interaction mimics any DM signal on the detector (neutrino floor).
In Fig.~\ref{fig:SIs}, the yellow points represent the viable models where both DM particles are expected to have signals in future experiments (points above the DARWIN experiment and below the current limit of XENONnT for the two DM components).
Also, notice, that in this model, the neutrinos are Dirac particles and contribute to the number of relativistic degrees of freedom in the early Universe via the kinetic mixing parameter~\cite{Calle:2019mxn} (see Appendix~\ref{appendix:Delta_N_Effective}). In Fig.~\ref{fig:SIs}, the magenta region shows the models with $\Delta N_{\text{eff}} < 0.3$ (PLANCK+BAO combinations~\cite{Planck:2018vyg}). Also, we show the blue region where $\epsilon=0$. Notice that it is inside the magenta region and represents the models with zero contribution to the relativistic degrees of freedom at tree level. However, although kinetic mixing could be generated radiatively~\cite{Davoudiasl:2005ks}, we check that it is always negligible in our model.  Notice that the magenta region outside in Fig. 9 is only excluded in the standard cosmological scenario. However, in non-standard cosmological scenarios, the $\Delta N_{eff}$ is relaxed as shown in Ref.~\cite{Biswas:2022fga}

Finally, we also checked the spin-dependent (SD) WIMP-neutron cross-section with \texttt{micrOMEGAs 6.0.3}. We found that all models have a SD cross-section $\sigma^{\text{SD}}\leq 10^{-43}$ cm$^2$, that is below the experimental constraints of XENON1T~\cite{XENON:2019rxp}, LZ~\cite{LZ:2022lsv} and its prospects as DARWIN~\cite{DARWIN:2016hyl}.

\begin{figure}[h]
 \centering
\includegraphics[scale=0.6]{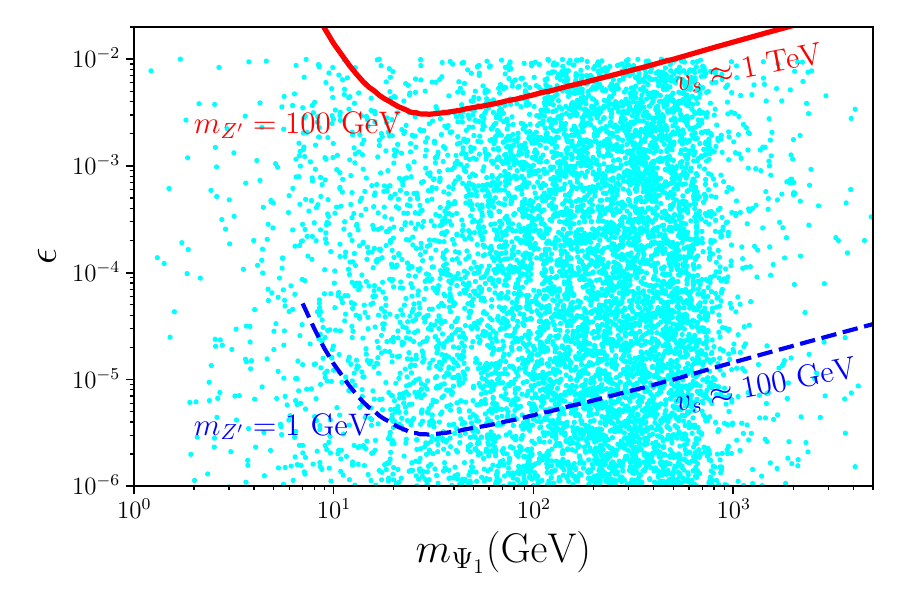}
 \caption{XENONnT contours limits for some dark photon masses projected in the plane of the kinetic mixing vs DM mass $m_{\Psi_1}$.}
\label{fig:epsilon-vs-DM}
\end{figure}

Fig.~\ref{fig:epsilon-vs-DM} studies the impact of the kinetic mixing parameter in the direct detection of DM. According to eq.~\eqref{eq:SI-vector}, the vector SI cross-section is a function of kinetic mixing, the DM mass, and the new $U(1)_D$ gauge coupling $g_D$. 
Moreover, Fig.~\ref{fig:epsilon-vs-DM} shows some contours for the XENONnT current limit for some fixed values of the dark photon $m_{Z'}$ and gauge coupling $g_D$ through the VEV $v_s$.
Notice that low masses of the dark photon prefer low values for the kinetic mixing parameter (blue dashed line). Also, kinetic mixing values below $10^{-6}$ are not restricted by XENONnT current limit because the SI cross-section is always under the neutrino floor. 

On the other hand, concerning the indirect detection of DM in this model, we performed an exhaustive numerical analysis with the \texttt{micrOMEGAs} program to compute today's annihilation rate. We realized that the principal annihilation channels for $\Psi_1$ are (also apply for $\Psi_2^1$): $\overline{\Psi}_1\Psi_1 \to\overline{\nu}_i\nu_i$, $\overline{\Psi}_1\Psi_1 \to Z^{\prime}Z^{\prime}$,  with a thermal average cross section $\langle \sigma v\rangle$ a little below the canonical value $\sim 3\times 10^{-26}\, \text{cm}^3/\text{s}$. However, the process involving the $Z^{\prime}$ triggers secondary particles in a 4-step cascade annihilation~\cite{Mardon:2009rc} where the unstable dark photon $Z^{\prime}$ decays in two states, principally in neutrinos $Z^{\prime}\to \overline{\nu}_i\nu_i$. 
This process could enhance the neutrino flux. However, this is beyond the work's scope.
We also have the annihilation channel $\overline{\Psi}_1\Psi_1 \to h_2Z^{\prime}$ where the extra scalar $h_2$ (Higgs-like) could decay in a leptonic or hadronic way with a significant modification of the gamma-ray spectrum, as analyzed in Ref.~\cite{Siqueira:2021lqj}. 
Finally, we also have direct annihilation channels (prompt channels) into two leptons $\overline{\Psi_1}\Psi_1\to \overline{e}_i e_i$ and two light quarks $\overline{\Psi}_1\Psi_1\to \overline{q}_iq_i$. However, in this model, these channels are suppressed by the kinetic mixing. After the scan shown in Tab.~\ref{tab:scan}, the $\langle \sigma v\rangle$ in these primary channels is under the value of $\sim 10^{-30}\, \text{cm}^3/\text{s}$, which is almost three-four orders of magnitude below the current limits and prospects of Fermi-LAT~\cite{Fermi-LAT:2015att} and the Cherenkov telescope arrays such as H.E.S.S~\cite{HESS:2016mib}, CTA~\cite{CTA:2020qlo}, SWGO~\cite{Schoorlemmer:2019gee} (see  Ref.~\cite{Siqueira:2021lqj} for more details).

Finally, notice that the DM particle could be captured by stars like our Sun and subsequently annihilated to a long-lived mediator. If the dark mediator escapes the Sun, it could produce gamma ray signals that could be detectable by experiments like the HAWC Observatory. However, for a dark photon mediator, the capture-annihilation equilibrium of the DM particle in the Sun is not obtained. The time scale $\tau_{eq}$ is greater than the age of the Sun~\cite{Niblaeus:2019gjk,Bell:2021pyy}. Also, we remark that a long-lived dark photon in this model has problems with BBN~\cite{Pospelov:2007mp}.

During the completion of this work, a similar work on chiral DM and radiative neutrino masses
from a gauged $U(1)$ symmetry appears~\cite{Babu:2024zoe}. 
There, with an alternative method to find some solutions, they explored a subset where all chiral fermions acquire masses after SSB. Without massless right-handed neutrinos, they generate scotogenic Majorana neutrino masses, and in our limit of a decoupled inert scalar sector, they need the kinetic mixing to explain the full DM phenomenology. 

\section{Conclusions}
 \label{sec:Conclusions}

This paper shows a complete UV realization of a secluded WIMP dark matter model with an extra Abelian gauge symmetry that includes two-component dark matter candidates, where Dirac neutrino masses are generated at one-loop via a scotogenic realization of the effective operator for Dirac neutrino masses in the SM. Our paper explains the relic abundance of dark matter, even without kinetic mixing. Also, it can be tested in direct detection experiments like DARWIN. However, the annihilation of dark matter particles into gamma rays is highly suppressed due to the dark nature of the Abelian gauge symmetry. The model's parameter space explains the relic abundance of DM and Dirac neutrino masses. It is also compatible with cosmological and theoretical constraints, including the branching ratio of SM into invisible, BBN restrictions, and the number of relativistic degrees of freedom in the early Universe, even without kinetic mixing.

\section{Acknowledgments}

\label{sec:Acknowledgments}

We thank Walter Tangarife for their enlightening discussions and reading of the manuscript. This work have been supported by Sostenibilidad UdeA, UdeA/CODI Grants 2022-52380 and 2023-59130, Minciencias Grants CD 82315 CT ICETEX 2021-1080.

\newpage
\appendix

\section{Dirac Neutrino Masses} 
\label{sec:Dirac-neutrino-masses}

In this model, we have a scotogenic realization of an effective operator for Dirac neutrino masses~\cite{Restrepo:2021kpq}:
\begin{align} 
\label{eq:nmo56}
    \mathcal{L}_{\text{eff}} = Y_{\nu}^{\alpha i} \, \left( \nu_{R\alpha}\right)^{\dagger} \,  \, L_i \cdot \, \tilde{H}
     \left(\frac{S^*}{\Lambda}\right)^\delta + \text{h.c.}\,,
\end{align}
where $\delta=1$, $(\cdot)$ is the $SU(2)_L$ dot product, $\tilde{H}=(0,-\frac{1}{\sqrt{2}}(v+h))^T$ and $\Lambda$ is the energy scale where Dirac mass terms for neutrinos are generated.
After spontaneous symmetry breaking, the mass Lagrangian for fermions in the dark sector that have two generations is:
\begin{align}
\mathcal{L}=\frac{v_s}{\sqrt{2}}\begin{pmatrix}
 \psi^1_L   &   \psi^2_L
\end{pmatrix}
\begin{pmatrix}
y^{11}_x & y^{12}_x \\
y^{21}_x & y^{22}_x
\end{pmatrix}
\begin{pmatrix}
(\psi^1_R)^{\dagger} \\
(\psi^1_R)^{\dagger}
\end{pmatrix}+ \text{h.c.}=\psi^T_Lm_{\psi}(\psi_R)^{\dagger}+ \text{h.c.}\,,
\end{align}
where the matrix $m_{\psi}$ is diagonalized as:
\begin{align}
Z_Lm_{\psi_2}Z_R^{\dagger}=(m_{\psi})^{\text{diag}}=\text{diag}(m_{\Psi_2^1},m_{\Psi_2^2})
\end{align}
with
\begin{equation}
\label{eq:ZLR}
Z_{L,R}=\begin{pmatrix}
\cos\theta_{L,R} & \sin\theta_{L,R} \\
-\sin\theta_{L,R} & \cos\theta_{L,R} 
\end{pmatrix}\,.
\end{equation}

We realize an effective Dirac neutrino mass operator via a scotogenic mechanism at one loop (see Feymann diagram~\ref{fig:DiracScotogenic}). 
At one loop level, we obtain a neutrino mass matrix:
\begin{equation}
\label{eq:righthandedneutrinomassterm}
(\mathcal{M}_{\nu})^{\alpha\beta}=\frac{1}{(4\pi)^2}\sum_{k=1}^2\sum_{l=1}^2\sum_{r=1}^2(y_{nL})^{\alpha k}(y_{nR})^{\beta r}(Z_R)^{kl}(Z_L)^{rl}M_l[U^{11}_{\Xi}f(m_1,M_l)U^{21}_{\Xi}+U^{12}_{\Xi}f(m_2,M_l)U^{22}_{\Xi}]\,,
\end{equation}
which can be written as
\begin{equation}
(\mathcal{M}_{\nu})^{\alpha\beta}=\sum_{k=1}^2\sum_{r=1}^2(y_{nL})^{\alpha k}f^{kr}(y_{nR})^{\beta r}\,,
\end{equation}
where
\begin{equation}
f^{kr}=\frac{1}{(4\pi)^2}\sum_{l=1}^2Z^{kl}_RZ^{rl}_LM_l[U^{11}_{\Xi}f(m_1,M_l)U^{21}_{\Xi}+U^{12}_{\Xi}f(m_2,M_l)U^{22}_{\Xi}]\,,
\end{equation}
\begin{equation}
f(m_i,M_l)=\frac{m^2_i\ln(m^2_i)-M^2_l\ln(M^2_l)}{m^2_i-M^2_l}\,.  
\end{equation}
We rotate to mass eigenstates from flavor eigenstates by using a biunitary transformation
\begin{equation}
M^\nu_\text{diag}=U^\dagger M^\nu V\,,
\end{equation}
where $U$ is the neutrino mixing matrix PMNS~\cite{Maki:1962mu}, $V$ is a unitary transformation matrix for right-handed neutrinos. The eigenvalues for neutrino masses can be related to data from PDG~\cite{PhysRevD.110.030001} (we consider normal hierarchy)
\begin{align}
\label{eq:defmnu1}
m_{\nu_{1}}\equiv & {0}\,,\\
\label{eq:defmnu2}
m_{\nu_{2}}\equiv & \sqrt{\Delta{m}^2_{21}}=8.678\times{10}^{-12}\ \text{GeV}\,,\\
\label{eq:defmnu3}
m_{\nu_{3}}\equiv & {m}_{\nu_2}+\sqrt{\Delta{m}^2_{23}}=5.01\times{10}^{-11}\ \text{GeV}\,.
\end{align}

In this paper, we consider the parameters $(y_{nL})^{\alpha k}$ as free parameters (We consider $(y_{nL})^{11}=0$ and $(y_{nL})^{12}=0$). Then, we write the couplings $y_{nR}$ in terms of neutrino observables
\begin{align}
(y_{nR})^{i1}=-\frac{(y_{nL})^{32}m_{\nu_{2}}U^{i2}-(y_{nL})^{22}m_{\nu_{3}}U^{i3}}{[(y_{nL})^{22}(y_{nL})^{31}-(y_{nL})^{21}(y_{nL})^{32}]\Lambda_1}\,,\\   
(y_{nR})^{i2}=-\frac{(y_{nL})^{31}m_{\nu_{2}}U^{i2}-(y_{nL})^{21}m_{\nu_{3}}U^{i3}}{[(y_{nL})^{22}(y_{nL})^{31}-(y_{nL})^{21}(y_{nL})^{32}]\Lambda_2}\,,
\end{align}
where
\begin{equation}
\Lambda_j=Z^{j1}_RZ^{j1}_L\mathcal{F}^1+Z^{j2}_RZ^{j2}_L\mathcal{F}^2\,,\\
\end{equation}
\begin{equation}
\mathcal{F}^k=m_{\Psi^k_2}(Z_N^{11}f(m_1,m_{\Psi^k_2})Z_N^{21}+Z_N^{12}f(m_2,m_{\Psi^k_2})Z_N^{22})\,,\\
\end{equation}
for $i=1,2,3$ and $j,k=1,2$.

\section{Diagonalization}
\label{sec:diagonalization-higgs}

After EWSB and SSB, we solve the tadpole equations and obtain the mass matrix
\begin{equation}
\label{eq:massmatrix}
 m_h^2 =\left(
\begin{array}{cc}
 \lambda _1 v^2 & \lambda _6 v v_s \\
 \lambda _6 v v_s & \lambda _5 v^2_s \\
\end{array}
\right)\,,
\end{equation}
that is diagonalized via an orthogonal matrix:
\begin{equation}
U=\begin{pmatrix}
\cos\theta & \sin\theta\\
-\sin\theta & \cos\theta
\end{pmatrix}\,,
\end{equation}
where $U^TMU=M_{\text{diagonal}}$
and
 \begin{equation}
M_{\text{diagonal}}=\left(
\begin{array}{cc}
 m^2_{h_1} & 0 \\
 0 & m^2_{h_2} \\
\end{array}
\right)\,.
\end{equation}
We can write the $\lambda_1$, $\lambda_5$, and $\lambda_6$ couplings can be written in terms of the mixing angle $\theta$ and the eigenvalues $m_{h_{1,2}}$ as
\begin{align}
\lambda_1= & \dfrac{1}{2v^2}\left(m_{h_1}^2+m_{h_2}^2-\dfrac{(m_{h_2}^2-m_{h_1}^2)}{\sqrt{1+\tan^2(2\theta)}}\right)\,,\\
    \lambda_5= & \frac{1}{2v^2_s}\left(m_{h_1}^2+m_{h_2}^2+\dfrac{(m_{h_2}^2-m_{h_1}^2)}{\sqrt{1+\tan^2(2\theta)}}\right)\,,\\
\lambda_6= & \frac{1}{2vv_s}\frac{m_{h_2}^2-m_{h_1}^2}{\sqrt{1+\tan^2(2\theta)}}\tan(2\theta)\,.
\end{align}%

\section{General thermal annihilation cross-section}
\label{sec:general-sv}

The thermally averaged cross section for DM annihilation for the process $\Psi\overline{\Psi} \to Z'Z'$ is written as powers of the relative velocity $v$ as
$\langle \sigma v\rangle = (a + b\, v^2 + O(v^4) ) $, where

\begin{align}
\label{eq:a-general}
a=\frac{g_D^4 \left(1-r^2\right)^{3/2} \left(Q_A^4 r^2+2 Q_A^2 Q_V^2 \left(4-3 r^2\right)+Q_V^4 r^2\right)}{4 \pi  r^2
   \left(r^2-2\right)^2 M_{\Psi }^2}\,,
\end{align}

\begin{align}
\label{eq:b-general}
    b\approx & \frac{g_D^4}{384 \pi m_{\Psi}^2  \left(m_h^2 -4 m_{\Psi }^2\right){}^2}\bigg[\nonumber\\
    &-\frac{8 \left(-8 \left(4 Q_A^2+81\right) Q_V^2 m_h^2 m_{\Psi }^2+4 Q_A^2 Q_V^2 m_h^4+m_{\Psi }^4 \left(32 \left(2
   Q_A^2+81\right) Q_V^2+59049\right)\right)}{r^2}\nonumber\\
   &+r^2 (-4 m_h^2 m_{\Psi }^2 \left(32 Q_A^4+Q_A^2 \left(64
   Q_V^2+81\right)+Q_V^2 \left(32 Q_V^2+81\right)\right))\nonumber\\
   &+r^2(16 m_h^4 \left(Q_A^2+Q_V^2\right)^2)\nonumber\\
   &+r^2(m_{\Psi }^4 \left(256
   Q_A^4+16 Q_A^2 \left(32 Q_V^2+81\right)+256 Q_V^4+1296 Q_V^2-98415\right))\nonumber \\
   &+2 (-8 m_h^2 m_{\Psi }^2
   \left(3 Q_A^4+6 Q_A^2 \left(7 Q_V^2-27\right)-7 Q_V^4+162 Q_V^2\right))\nonumber\\
   &+2(m_h^4 \left(3 Q_A^4+42 Q_A^2
   Q_V^2-7 Q_V^4\right))\nonumber\\
   &+2(m_{\Psi }^4 \left(48 Q_A^4+96 Q_A^2 \left(7 Q_V^2-54\right)-112 Q_V^4+5184
   Q_V^2+177147\right) )\nonumber\\
   &+\frac{16 \left(-4 Q_V^2 \left(4 Q_V^2+81\right) m_h^2 m_{\Psi }^2+2 Q_V^4 m_h^4+\left(32
   Q_V^4+1296 Q_V^2+19683\right) m_{\Psi }^4\right)}{r^4}\bigg]\,,
\end{align}
with $r=m_{Z'}/m_{\Psi}$, $Q_A=(q_R-q_L)/2$ and  $Q_V=(q_R+q_L)/2$ are the axial-vector and vector charges.
$q_L, q_R$ are the left and right Weyl fermion charges that form the Dirac fermion $\Psi$.
The s-wave matches the expression in Refs.\cite{Bell:2016uhg, Alves:2015pea,Alves:2015mua} and the p-wave is reported for the first time. Although those expressions are general, in this work, the lightest DM particle is $\Psi_{1}$: with $q_L=4, q_R=5$ as is shown in Tab.~\ref{tab:pickedsltn}, and therefore, the s-wave and the p-wave contributions become:
\begin{align}
\label{eq:a-Model}
    a = \frac{g_D^4 \left(1-r^2\right)^{3/2} \left(1519\, r^2+162\right)}{16 \pi  m_{\Psi}^2 r^2 \left(r^2-2\right)^2}\,,
\end{align}
\begin{align}
\label{eq:b-Model}
    b
    & \approx \frac{g_D^4}{384 \pi m_{\Psi }^2 \left(m_h^2 -4 m_{\Psi }^2\right){}^2} \bigg[
    \frac{16 \left(-13122 m_h^2 m_{\Psi }^2+\frac{6561 m_h^4}{8}+59049 m_{\Psi }^4\right)}{r^4}\nonumber \\
    &-\frac{8 \left(-13284 m_h^2 m_{\Psi }^2+\frac{81 m_h^4}{4}+111861 m_{\Psi }^4\right)}{r^2} +r^2 \left(-60434 m_h^2 m_{\Psi }^2+6724 m_h^4+35737
   m_{\Psi }^4\right)\nonumber\\
   &+2 \left(-4659 m_h^2 m_{\Psi
   }^2-\frac{21261 m_h^4}{8}+238305 m_{\Psi }^4\right) \bigg] \,.
\end{align}
Note that in the vector-like limit $q_L=q_R=1$, the Eqs.~\eqref{eq:a-general}, ~\eqref{eq:b-general} give
\begin{align}
    a &=\frac{g_D^4 \left(1-r^2\right)^{3/2}}{4 \pi  m_{\Psi}^2 \left(r^2-2\right)^2}\nonumber \\
    b &\approx 0\,.
\end{align}
Those equations match the ones in Refs.~\cite{Ma:2021szi, Ma:2022uhi}.

\section{Higgs Decays}
\label{appendix:Higgs_Decays}

\subsection{Higgs decay into dark fermions}

If $m_{\Psi_i}< m_{h_1}/2=64.5\, \text{GeV}$, the Higgs $h_1$ decays into DM particles $\Psi_i=\Psi_1,\Psi^2_1$ through the s-channel with the $Z^{\prime}$ boson with a decay width given by
\begin{equation}
\Gamma(h_1\to\overline{\Psi}_i\Psi_i)=\frac{m_{h_1}|y_i|^2\sin^2(\theta)}{16\pi}\left(1-\frac{4}{r_i}\right)^{\frac{3}{2}}
\end{equation}
where $r_i=m^2_{h_1}/m_{\Psi_i^2}$ and
\begin{align}
y_1= & y_c \nonumber \\
y_2= &(Z_R)^{11}[(Z_L)^{11}(y_{x})^{11}+(Z_L)^{12}(y_{x})^{12}]+(Z_R)^{12}[(Z_L)^{11}(y_{x})^{12}+(Z_L)^{12}(y_{x})^{22}]/\sqrt{2}\,.
\end{align}

\section{$Z$ decay into dark fermions}
\label{appendix:Z_Decays}

If $m_{\Psi_k}<m_Z/2$ the decay of $Z$ boson into dark fermions is kinematically allowed, the decay width is:
\begin{equation}
\Gamma(Z\to\overline{\Psi}_k\Psi_k)=\frac{g_1^2\epsilon^2\operatorname{s}^2_W}{24\pi}\sqrt{1-\frac{4}{r}}\left[(a_k^2+b^2_k)\left(1-\frac{1}{r}\right)+6a_kb_k\frac{1}{r}\right]\,,
\end{equation}
where $a_k=(q_{\psi^k_L}-q_{\psi^k_R})$, $b_k=(q_{\psi^k_L}+q_{\psi^k_R})$, $\operatorname{s}_W$ is the sine of the Weinberg angle, and $r= m^2_Z/m^2_{\Psi_k}$.

\section{Contribution to the number of relativistic degrees of freedom in the early Universe}
\label{appendix:Delta_N_Effective}

Models where neutrinos are Dirac fermions can have additional contributions to the number of relativistic degrees of freedom in the early Universe~\cite{Calle:2019mxn}. Right-handed neutrinos can be in thermal equilibrium with SM plasma for non-zero kinetic mixing. To maintain equilibrium, the following condition holds:
\begin{equation}
\Gamma_{\nu_{R}}(T) > 3 H\,n_{\nu_R} \,,   
\end{equation}
where $H$ is the Hubble rate,  $n_{\nu_R}$ is the number density of right-handed neutrinos and $\Gamma$ is the interaction rate between right-handed neutrinos and SM plasma, which is given by:
\begin{equation}
\Gamma_{\nu_{R}}(T) = \frac{49 \pi^{5} T^{5}}{97200 \zeta(3)} \left( \frac{1}{M_{Z^{\prime}}} \right)^4(g^{\nu_R}_Z)^2\sum_{f} N^{C}_{f}[(g^{f_L}_Z)^2+(g^{f_R}_Z)^2]\,,
\end{equation}
where:
\begin{equation}
g^{\nu_R}_Z=9(g_D\operatorname{s}_{W'}+g_1\epsilon\operatorname{c}_{W'}\operatorname{s}_W)\,, 
\end{equation}
that can be approximated as $g^{\nu_R}_Z  \approx 9 g_1\epsilon\sin(\theta_{W})$ and $g^{f_L}_Z$ and $g^{f_R}_Z$ are the couplings of the SM fermions to the $Z$ boson. The contribution to the $\Delta N_{\text{eff}}$ is:
\begin{align}
\Delta N_{\text{eff}} = N_{\text{eff}} - N^{\text{SM}}_{\text{eff}} = N_{\nu_R} \left( \frac{T_{\nu_{R}}}{T_{\nu_{L}}} \right)^{4} = N_{\nu_R} \left( \frac{g(T^{\nu_{L}}_{\text{dec}})}{g(T^{\nu_{R}}_{\text{dec}})} \right)^{4/3}\,.
\end{align}
where $T^{\nu_{R}}_{\text{dec}}$ is the temperature of the decoupling of the right-handed neutrinos from SM plasma and $T_{\nu_{L}}$ is the temperature of decoupling of left-handed neutrinos from thermal plasma $T_{\nu_{L}}\,\sim\,{2.3}\,\text{MeV}$.

\bibliographystyle{apsrev4-1long}
\bibliography{susy}

\begin{thebibliography}{10}%
\makeatletter
\providecommand \@ifxundefined [1]{%
 \ifx #1\undefined \expandafter \@firstoftwo
 \else \expandafter \@secondoftwo
\fi
}%
\providecommand \@ifnum [1]{%
 \ifnum #1\expandafter \@firstoftwo
 \else \expandafter \@secondoftwo
\fi
}%
\providecommand \enquote [1]{``#1''}%
\providecommand \bibnamefont  [1]{#1}%
\providecommand \bibfnamefont [1]{#1}%
\providecommand \citenamefont [1]{#1}%
\providecommand\href[0]{\@sanitize\@href}%
\providecommand\@href[1]{\endgroup\@@startlink{#1}\endgroup\@@href}%
\providecommand\@@href[1]{#1\@@endlink}%
\providecommand \@sanitize [0]{\begingroup\catcode`\&12\catcode`\#12\relax}%
\@ifxundefined \pdfoutput {\@firstoftwo}{%
 \@ifnum{\z@=\pdfoutput}{\@firstoftwo}{\@secondoftwo}%
}{%
 \providecommand\@@startlink[1]{\leavevmode\special{html:<a href="#1">}}%
 \providecommand\@@endlink[0]{\special{html:</a>}}%
}{%
 \providecommand\@@startlink[1]{%
  \leavevmode
  \pdfstartlink
   attr{/Border[0 0 1 ]/H/I/C[0 1 1]}%
   user{/Subtype/Link/A<</Type/Action/S/URI/URI(#1)>>}%
  \relax
 }%
 \providecommand\@@endlink[0]{\pdfendlink}%
}%
\providecommand \url  [0]{\begingroup\@sanitize \@url }%
\providecommand \@url [1]{\endgroup\@href {#1}{\urlprefix}}%
\providecommand \urlprefix [0]{URL }%
\providecommand \Eprint[0]{\href }%
\@ifxundefined \urlstyle {%
  \providecommand \doi [1]{doi:\discretionary{}{}{}#1}%
}{%
  \providecommand \doi [0]{doi:\discretionary{}{}{}\begingroup
  \urlstyle{rm}\Url }%
}%
\providecommand \doibase [0]{http://dx.doi.org/}%
\providecommand \Doi[1]{\href{\doibase#1}}%
\providecommand \bibAnnote [3]{%
  \BibitemShut{#1}%
  \begin{quotation}\noindent
    \textsc{Key:}\ #2\\\textsc{Annotation:}\ #3%
  \end{quotation}%
}%
\providecommand \bibAnnoteFile [2]{%
  \IfFileExists{#2}{\bibAnnote {#1} {#2} {\input{#2}}}{}%
}%
\providecommand \typeout [0]{\immediate \write \m@ne }%
\providecommand \selectlanguage [0]{\@gobble}%
\providecommand \bibinfo [0]{\@secondoftwo}%
\providecommand \bibfield [0]{\@secondoftwo}%
\providecommand \translation [1]{[#1]}%
\providecommand \BibitemOpen[0]{}%
\providecommand \bibitemStop [0]{}%
\providecommand \bibitemNoStop [0]{.\EOS\space}%
\providecommand \EOS [0]{\spacefactor3000\relax}%
\providecommand \BibitemShut [1]{\csname bibitem#1\endcsname}%
\bibitem{Steigman:1984ac}%
  \BibitemOpen
  \bibfield{author}{%
  \bibinfo {author} {\bibfnamefont{Gary}\ \bibnamefont{Steigman}}\ and\
  \bibinfo {author} {\bibfnamefont{Michael~S.}\ \bibnamefont{Turner}},\ }%
  \bibfield{title}{%
  \enquote{\bibinfo {title} {{Cosmological Constraints on the Properties of
  Weakly Interacting Massive Particles}},}\ }%
  \bibfield{journal}{%
  \Doi{10.1016/0550-3213(85)90537-1}{\bibinfo {journal} {Nucl. Phys. B}}\ }%
  \textbf{\bibinfo {volume} {253}},\ \bibinfo {pages} {375--386} (\bibinfo
  {year} {1985})%
  \bibAnnoteFile{NoStop}{Steigman:1984ac}%
\bibitem{Cushman:2013zza}%
  \BibitemOpen
  \bibfield{author}{%
  \bibinfo {author} {\bibfnamefont{P.}~\bibnamefont{Cushman}} \emph{et~al.},\
  }%
  \enquote{\bibinfo {title} {{Working Group Report: WIMP Dark Matter Direct
  Detection}},}\ in\ \emph{\bibinfo {booktitle} {{Snowmass 2013}: {Snowmass on
  the Mississippi}}}\ (\bibinfo {year} {2013})\
  \Eprint{http://arxiv.org/abs/1310.8327}{arXiv:1310.8327 [hep-ex]}%
  \bibAnnoteFile{NoStop}{Cushman:2013zza}%
\bibitem{Bertone:2016nfn}%
  \BibitemOpen
  \bibfield{author}{%
  \bibinfo {author} {\bibfnamefont{Gianfranco}\ \bibnamefont{Bertone}}\ and\
  \bibinfo {author} {\bibfnamefont{Dan}\ \bibnamefont{Hooper}},\ }%
  \bibfield{title}{%
  \enquote{\bibinfo {title} {{History of dark matter}},}\ }%
  \bibfield{journal}{%
  \Doi{10.1103/RevModPhys.90.045002}{\bibinfo {journal} {Rev. Mod. Phys.}}\ }%
  \textbf{\bibinfo {volume} {90}},\ \bibinfo {pages} {045002} (\bibinfo {year}
  {2018}),\ \Eprint{http://arxiv.org/abs/1605.04909}{arXiv:1605.04909
  [astro-ph.CO]}%
  \bibAnnoteFile{NoStop}{Bertone:2016nfn}%
\bibitem{Feng:2022rxt}%
  \BibitemOpen
  \bibfield{author}{%
  \bibinfo {author} {\bibfnamefont{Jonathan~L.}\ \bibnamefont{Feng}},\ }%
  \bibfield{title}{%
  \enquote{\bibinfo {title} {{The WIMP paradigm: Theme and variations}},}\ }%
  \bibfield{journal}{%
  \Doi{10.21468/SciPostPhysLectNotes.71}{\bibinfo {journal} {SciPost Phys.
  Lect. Notes}}\ }%
  \textbf{\bibinfo {volume} {71}},\ \bibinfo {pages} {1} (\bibinfo {year}
  {2023}),\ \Eprint{http://arxiv.org/abs/2212.02479}{arXiv:2212.02479
  [hep-ph]}%
  \bibAnnoteFile{NoStop}{Feng:2022rxt}%
\bibitem{Preskill:1982cy}%
  \BibitemOpen
  \bibfield{author}{%
  \bibinfo {author} {\bibfnamefont{John}\ \bibnamefont{Preskill}}, \bibinfo
  {author} {\bibfnamefont{Mark~B.}\ \bibnamefont{Wise}},\ and\ \bibinfo
  {author} {\bibfnamefont{Frank}\ \bibnamefont{Wilczek}},\ }%
  \bibfield{title}{%
  \enquote{\bibinfo {title} {{Cosmology of the Invisible Axion}},}\ }%
  \bibfield{journal}{%
  \Doi{10.1016/0370-2693(83)90637-8}{\bibinfo {journal} {Phys. Lett. B}}\ }%
  \textbf{\bibinfo {volume} {120}},\ \bibinfo {pages} {127--132} (\bibinfo
  {year} {1983})%
  \bibAnnoteFile{NoStop}{Preskill:1982cy}%
\bibitem{Abbott:1982af}%
  \BibitemOpen
  \bibfield{author}{%
  \bibinfo {author} {\bibfnamefont{L.~F.}\ \bibnamefont{Abbott}}\ and\ \bibinfo
  {author} {\bibfnamefont{P.}~\bibnamefont{Sikivie}},\ }%
  \bibfield{title}{%
  \enquote{\bibinfo {title} {{A Cosmological Bound on the Invisible Axion}},}\
  }%
  \bibfield{journal}{%
  \Doi{10.1016/0370-2693(83)90638-X}{\bibinfo {journal} {Phys. Lett. B}}\ }%
  \textbf{\bibinfo {volume} {120}},\ \bibinfo {pages} {133--136} (\bibinfo
  {year} {1983})%
  \bibAnnoteFile{NoStop}{Abbott:1982af}%
\bibitem{Dine:1982ah}%
  \BibitemOpen
  \bibfield{author}{%
  \bibinfo {author} {\bibfnamefont{Michael}\ \bibnamefont{Dine}}\ and\ \bibinfo
  {author} {\bibfnamefont{Willy}\ \bibnamefont{Fischler}},\ }%
  \bibfield{title}{%
  \enquote{\bibinfo {title} {{The Not So Harmless Axion}},}\ }%
  \bibfield{journal}{%
  \Doi{10.1016/0370-2693(83)90639-1}{\bibinfo {journal} {Phys. Lett. B}}\ }%
  \textbf{\bibinfo {volume} {120}},\ \bibinfo {pages} {137--141} (\bibinfo
  {year} {1983})%
  \bibAnnoteFile{NoStop}{Dine:1982ah}%
\bibitem{Kopp:2021jlk}%
  \BibitemOpen
  \bibfield{author}{%
  \bibinfo {author} {\bibfnamefont{Joachim}\ \bibnamefont{Kopp}},\ }%
  \bibfield{title}{%
  \enquote{\bibinfo {title} {{Sterile neutrinos as dark matter candidates}},}\
  }%
  \bibfield{journal}{%
  \Doi{10.21468/SciPostPhysLectNotes.36}{\bibinfo {journal} {SciPost Phys.
  Lect. Notes}}\ }%
  \textbf{\bibinfo {volume} {36}},\ \bibinfo {pages} {1} (\bibinfo {year}
  {2022}),\ \Eprint{http://arxiv.org/abs/2109.00767}{arXiv:2109.00767
  [hep-ph]}%
  \bibAnnoteFile{NoStop}{Kopp:2021jlk}%
\bibitem{Carr:2020gox}%
  \BibitemOpen
  \bibfield{author}{%
  \bibinfo {author} {\bibfnamefont{Bernard}\ \bibnamefont{Carr}}, \bibinfo
  {author} {\bibfnamefont{Kazunori}\ \bibnamefont{Kohri}}, \bibinfo {author}
  {\bibfnamefont{Yuuiti}\ \bibnamefont{Sendouda}},\ and\ \bibinfo {author}
  {\bibfnamefont{Jun'ichi}\ \bibnamefont{Yokoyama}},\ }%
  \bibfield{title}{%
  \enquote{\bibinfo {title} {{Constraints on primordial black holes}},}\ }%
  \bibfield{journal}{%
  \Doi{10.1088/1361-6633/ac1e31}{\bibinfo {journal} {Rept. Prog. Phys.}}\ }%
  \textbf{\bibinfo {volume} {84}},\ \bibinfo {pages} {116902} (\bibinfo {year}
  {2021}),\ \Eprint{http://arxiv.org/abs/2002.12778}{arXiv:2002.12778
  [astro-ph.CO]}%
  \bibAnnoteFile{NoStop}{Carr:2020gox}%
\bibitem{Bird:2016dcv}%
  \BibitemOpen
  \bibfield{author}{%
  \bibinfo {author} {\bibfnamefont{Simeon}\ \bibnamefont{Bird}}, \bibinfo
  {author} {\bibfnamefont{Ilias}\ \bibnamefont{Cholis}}, \bibinfo {author}
  {\bibfnamefont{Julian~B.}\ \bibnamefont{Mu\~noz}}, \bibinfo {author}
  {\bibfnamefont{Yacine}\ \bibnamefont{Ali-Ha\"\i{}moud}}, \bibinfo {author}
  {\bibfnamefont{Marc}\ \bibnamefont{Kamionkowski}}, \bibinfo {author}
  {\bibfnamefont{Ely~D.}\ \bibnamefont{Kovetz}}, \bibinfo {author}
  {\bibfnamefont{Alvise}\ \bibnamefont{Raccanelli}},\ and\ \bibinfo {author}
  {\bibfnamefont{Adam~G.}\ \bibnamefont{Riess}},\ }%
  \bibfield{title}{%
  \enquote{\bibinfo {title} {{Did LIGO detect dark matter?}}.}\ }%
  \bibfield{journal}{%
  \Doi{10.1103/PhysRevLett.116.201301}{\bibinfo {journal} {Phys. Rev. Lett.}}\
  }%
  \textbf{\bibinfo {volume} {116}},\ \bibinfo {pages} {201301} (\bibinfo {year}
  {2016}),\ \Eprint{http://arxiv.org/abs/1603.00464}{arXiv:1603.00464
  [astro-ph.CO]}%
  \bibAnnoteFile{NoStop}{Bird:2016dcv}%
\bibitem{Barman:2024kfj}%
  \BibitemOpen
  \bibfield{author}{%
  \bibinfo {author} {\bibfnamefont{Basabendu}\ \bibnamefont{Barman}}, \bibinfo
  {author} {\bibfnamefont{Kousik}\ \bibnamefont{Loho}},\ and\ \bibinfo {author}
  {\bibfnamefont{\'Oscar}\ \bibnamefont{Zapata}},\ }%
  \bibfield{title}{%
  \enquote{\bibinfo {title} {{Asymmetries from a charged memory-burdened
  PBH}},}\ }%
  \bibfield{journal}{%
  \Doi{10.1088/1475-7516/2025/02/052}{\bibinfo {journal} {JCAP}}\ }%
  \textbf{\bibinfo {volume} {02}},\ \bibinfo {pages} {052} (\bibinfo {year}
  {2025}),\ \Eprint{http://arxiv.org/abs/2412.13254}{arXiv:2412.13254
  [hep-ph]}%
  \bibAnnoteFile{NoStop}{Barman:2024kfj}%
\bibitem{Chulia:2021jgv}%
  \BibitemOpen
  \bibfield{author}{%
  \bibinfo {author} {\bibfnamefont{Salvador~Centelles}\
  \bibnamefont{Chuli\'a}},\ }%
  \bibfield{title}{%
  \enquote{\bibinfo {title} {{Theory and phenomenology of Dirac neutrinos}},}\
  }%
   (\bibinfo {month} {10}\ \bibinfo {year} {2021}),\
  \Eprint{http://arxiv.org/abs/2110.15755}{arXiv:2110.15755 [hep-ph]}%
  \bibAnnoteFile{NoStop}{Chulia:2021jgv}%
\bibitem{Cai:2017jrq}%
  \BibitemOpen
  \bibfield{author}{%
  \bibinfo {author} {\bibfnamefont{Yi}~\bibnamefont{Cai}}, \bibinfo {author}
  {\bibfnamefont{Juan}\ \bibnamefont{Herrero-Garc\'\i{}a}}, \bibinfo {author}
  {\bibfnamefont{Michael~A.}\ \bibnamefont{Schmidt}}, \bibinfo {author}
  {\bibfnamefont{Avelino}\ \bibnamefont{Vicente}},\ and\ \bibinfo {author}
  {\bibfnamefont{Raymond~R.}\ \bibnamefont{Volkas}},\ }%
  \bibfield{title}{%
  \enquote{\bibinfo {title} {{From the trees to the forest: a review of
  radiative neutrino mass models}},}\ }%
  \bibfield{journal}{%
  \Doi{10.3389/fphy.2017.00063}{\bibinfo {journal} {Front. in Phys.}}\ }%
  \textbf{\bibinfo {volume} {5}},\ \bibinfo {pages} {63} (\bibinfo {year}
  {2017}),\ \Eprint{http://arxiv.org/abs/1706.08524}{arXiv:1706.08524
  [hep-ph]}%
  \bibAnnoteFile{NoStop}{Cai:2017jrq}%
\bibitem{Pospelov:2007mp}%
  \BibitemOpen
  \bibfield{author}{%
  \bibinfo {author} {\bibfnamefont{Maxim}\ \bibnamefont{Pospelov}}, \bibinfo
  {author} {\bibfnamefont{Adam}\ \bibnamefont{Ritz}},\ and\ \bibinfo {author}
  {\bibfnamefont{Mikhail~B.}\ \bibnamefont{Voloshin}},\ }%
  \bibfield{title}{%
  \enquote{\bibinfo {title} {{Secluded WIMP Dark Matter}},}\ }%
  \bibfield{journal}{%
  \Doi{10.1016/j.physletb.2008.02.052}{\bibinfo {journal} {Phys. Lett. B}}\ }%
  \textbf{\bibinfo {volume} {662}},\ \bibinfo {pages} {53--61} (\bibinfo {year}
  {2008}),\ \Eprint{http://arxiv.org/abs/0711.4866}{arXiv:0711.4866 [hep-ph]}%
  \bibAnnoteFile{NoStop}{Pospelov:2007mp}%
\bibitem{Pospelov:2008zw}%
  \BibitemOpen
  \bibfield{author}{%
  \bibinfo {author} {\bibfnamefont{Maxim}\ \bibnamefont{Pospelov}},\ }%
  \bibfield{title}{%
  \enquote{\bibinfo {title} {{Secluded U(1) below the weak scale}},}\ }%
  \bibfield{journal}{%
  \Doi{10.1103/PhysRevD.80.095002}{\bibinfo {journal} {Phys. Rev. D}}\ }%
  \textbf{\bibinfo {volume} {80}},\ \bibinfo {pages} {095002} (\bibinfo {year}
  {2009}),\ \Eprint{http://arxiv.org/abs/0811.1030}{arXiv:0811.1030 [hep-ph]}%
  \bibAnnoteFile{NoStop}{Pospelov:2008zw}%
\bibitem{Holdom:1985ag}%
  \BibitemOpen
  \bibfield{author}{%
  \bibinfo {author} {\bibfnamefont{Bob}\ \bibnamefont{Holdom}},\ }%
  \bibfield{title}{%
  \enquote{\bibinfo {title} {{Two U(1)'s and Epsilon Charge Shifts}},}\ }%
  \bibfield{journal}{%
  \Doi{10.1016/0370-2693(86)91377-8}{\bibinfo {journal} {Phys. Lett. B}}\ }%
  \textbf{\bibinfo {volume} {166}},\ \bibinfo {pages} {196--198} (\bibinfo
  {year} {1986})%
  \bibAnnoteFile{NoStop}{Holdom:1985ag}%
\bibitem{Cline:2014dwa}%
  \BibitemOpen
  \bibfield{author}{%
  \bibinfo {author} {\bibfnamefont{James~M.}\ \bibnamefont{Cline}}, \bibinfo
  {author} {\bibfnamefont{Grace}\ \bibnamefont{Dupuis}}, \bibinfo {author}
  {\bibfnamefont{Zuowei}\ \bibnamefont{Liu}},\ and\ \bibinfo {author}
  {\bibfnamefont{Wei}\ \bibnamefont{Xue}},\ }%
  \bibfield{title}{%
  \enquote{\bibinfo {title} {{The windows for kinetically mixed Z'-mediated
  dark matter and the galactic center gamma ray excess}},}\ }%
  \bibfield{journal}{%
  \Doi{10.1007/JHEP08(2014)131}{\bibinfo {journal} {JHEP}}\ }%
  \textbf{\bibinfo {volume} {08}},\ \bibinfo {pages} {131} (\bibinfo {year}
  {2014}),\ \Eprint{http://arxiv.org/abs/1405.7691}{arXiv:1405.7691 [hep-ph]}%
  \bibAnnoteFile{NoStop}{Cline:2014dwa}%
\bibitem{Fabbrichesi:2020wbt}%
  \BibitemOpen
  \bibfield{author}{%
  \bibinfo {author} {\bibfnamefont{Marco}\ \bibnamefont{Fabbrichesi}}, \bibinfo
  {author} {\bibfnamefont{Emidio}\ \bibnamefont{Gabrielli}},\ and\ \bibinfo
  {author} {\bibfnamefont{Gaia}\ \bibnamefont{Lanfranchi}},\ }%
  \Doi{10.1007/978-3-030-62519-1}{\emph{\bibinfo {title} {{The Dark
  Photon}}}},\ SpringerBriefs in Physics\ (\bibinfo {publisher} {Springer},\
  \bibinfo {year} {2020})\ ISBN \bibinfo {isbn} {978-3-030--62519-1},\ pp.\
  \bibinfo {pages} {X,78},\
  \Eprint{http://arxiv.org/abs/2005.01515}{arXiv:2005.01515 [hep-ph]}%
  \bibAnnoteFile{NoStop}{Fabbrichesi:2020wbt}%
\bibitem{Bell:2016uhg}%
  \BibitemOpen
  \bibfield{author}{%
  \bibinfo {author} {\bibfnamefont{Nicole~F.}\ \bibnamefont{Bell}}, \bibinfo
  {author} {\bibfnamefont{Yi}~\bibnamefont{Cai}},\ and\ \bibinfo {author}
  {\bibfnamefont{Rebecca~K.}\ \bibnamefont{Leane}},\ }%
  \bibfield{title}{%
  \enquote{\bibinfo {title} {{Impact of mass generation for spin-1 mediator
  simplified models}},}\ }%
  \bibfield{journal}{%
  \Doi{10.1088/1475-7516/2017/01/039}{\bibinfo {journal} {JCAP}}\ }%
  \textbf{\bibinfo {volume} {01}},\ \bibinfo {pages} {039} (\bibinfo {year}
  {2017}),\ \Eprint{http://arxiv.org/abs/1610.03063}{arXiv:1610.03063
  [hep-ph]}%
  \bibAnnoteFile{NoStop}{Bell:2016uhg}%
\bibitem{Stueckelberg:1938hvi}%
  \BibitemOpen
  \bibfield{author}{%
  \bibinfo {author} {\bibfnamefont{E.~C.~G.}\ \bibnamefont{Stueckelberg}},\ }%
  \bibfield{title}{%
  \enquote{\bibinfo {title} {{Interaction energy in electrodynamics and in the
  field theory of nuclear forces}},}\ }%
  \bibfield{journal}{%
  \Doi{10.5169/seals-110852}{\bibinfo {journal} {Helv. Phys. Acta}}\ }%
  \textbf{\bibinfo {volume} {11}},\ \bibinfo {pages} {225--244} (\bibinfo
  {year} {1938})%
  \bibAnnoteFile{NoStop}{Stueckelberg:1938hvi}%
\bibitem{Hambye:2019dwd}%
  \BibitemOpen
  \bibfield{author}{%
  \bibinfo {author} {\bibfnamefont{Thomas}\ \bibnamefont{Hambye}}, \bibinfo
  {author} {\bibfnamefont{Michel H.~G.}\ \bibnamefont{Tytgat}}, \bibinfo
  {author} {\bibfnamefont{J\'er\^ome}\ \bibnamefont{Vandecasteele}},\ and\
  \bibinfo {author} {\bibfnamefont{Laurent}\ \bibnamefont{Vanderheyden}},\ }%
  \bibfield{title}{%
  \enquote{\bibinfo {title} {{Dark matter from dark photons: a taxonomy of dark
  matter production}},}\ }%
  \bibfield{journal}{%
  \Doi{10.1103/PhysRevD.100.095018}{\bibinfo {journal} {Phys. Rev. D}}\ }%
  \textbf{\bibinfo {volume} {100}},\ \bibinfo {pages} {095018} (\bibinfo {year}
  {2019}),\ \Eprint{http://arxiv.org/abs/1908.09864}{arXiv:1908.09864
  [hep-ph]}%
  \bibAnnoteFile{NoStop}{Hambye:2019dwd}%
\bibitem{Costa:2019zzy}%
  \BibitemOpen
  \bibfield{author}{%
  \bibinfo {author} {\bibfnamefont{Davi~B.}\ \bibnamefont{Costa}}, \bibinfo
  {author} {\bibfnamefont{Bogdan~A.}\ \bibnamefont{Dobrescu}},\ and\ \bibinfo
  {author} {\bibfnamefont{Patrick~J.}\ \bibnamefont{Fox}},\ }%
  \bibfield{title}{%
  \enquote{\bibinfo {title} {{General Solution to the U(1) Anomaly
  Equations}},}\ }%
  \bibfield{journal}{%
  \Doi{10.1103/PhysRevLett.123.151601}{\bibinfo {journal} {Phys. Rev. Lett.}}\
  }%
  \textbf{\bibinfo {volume} {123}},\ \bibinfo {pages} {151601} (\bibinfo {year}
  {2019}),\ \Eprint{http://arxiv.org/abs/1905.13729}{arXiv:1905.13729
  [hep-th]}%
  \bibAnnoteFile{NoStop}{Costa:2019zzy}%
\bibitem{Restrepo:2021kpq}%
  \BibitemOpen
  \bibfield{author}{%
  \bibinfo {author} {\bibfnamefont{Diego}\ \bibnamefont{Restrepo}}\ and\
  \bibinfo {author} {\bibfnamefont{David}\ \bibnamefont{Suarez}},\ }%
  \bibfield{title}{%
  \enquote{\bibinfo {title} {{Effective Dirac Neutrino Mass Operator in the
  Standard Model With a Local Abelian Extension}},}\ }%
  \bibfield{journal}{%
  \Doi{10.3389/fphy.2022.838531}{\bibinfo {journal} {Front. in Phys.}}\ }%
  \textbf{\bibinfo {volume} {10}},\ \bibinfo {pages} {838531} (\bibinfo {year}
  {2022}),\ \Eprint{http://arxiv.org/abs/2112.09524}{arXiv:2112.09524
  [hep-ph]}%
  \bibAnnoteFile{NoStop}{Restrepo:2021kpq}%
\bibitem{Wong:2020obo}%
  \BibitemOpen
  \bibfield{author}{%
  \bibinfo {author} {\bibfnamefont{Chi-Fong}\ \bibnamefont{Wong}},\ }%
  \bibfield{title}{%
  \enquote{\bibinfo {title} {{Anomaly-free chiral $U(1)_D$ and its scotogenic
  implication}},}\ }%
  \bibfield{journal}{%
  \Doi{10.1016/j.dark.2021.100818}{\bibinfo {journal} {Phys. Dark Univ.}}\ }%
  \textbf{\bibinfo {volume} {32}},\ \bibinfo {pages} {100818} (\bibinfo {year}
  {2021}),\ \Eprint{http://arxiv.org/abs/2008.08573}{arXiv:2008.08573
  [hep-ph]}%
  \bibAnnoteFile{NoStop}{Wong:2020obo}%
\bibitem{Bernal:2021ezl}%
  \BibitemOpen
  \bibfield{author}{%
  \bibinfo {author} {\bibfnamefont{Nicol\'as}\ \bibnamefont{Bernal}}, \bibinfo
  {author} {\bibfnamefont{Juli\'an}\ \bibnamefont{Calle}},\ and\ \bibinfo
  {author} {\bibfnamefont{Diego}\ \bibnamefont{Restrepo}},\ }%
  \bibfield{title}{%
  \enquote{\bibinfo {title} {{Anomaly-free Abelian gauge symmetries with Dirac
  scotogenic models}},}\ }%
  \bibfield{journal}{%
  \Doi{10.1103/PhysRevD.103.095032}{\bibinfo {journal} {Phys. Rev. D}}\ }%
  \textbf{\bibinfo {volume} {103}},\ \bibinfo {pages} {095032} (\bibinfo {year}
  {2021}),\ \Eprint{http://arxiv.org/abs/2102.06211}{arXiv:2102.06211
  [hep-ph]}%
  \bibAnnoteFile{NoStop}{Bernal:2021ezl}%
\bibitem{Babu:2003is}%
  \BibitemOpen
  \bibfield{author}{%
  \bibinfo {author} {\bibfnamefont{K.~S.}\ \bibnamefont{Babu}}\ and\ \bibinfo
  {author} {\bibfnamefont{Gerhart}\ \bibnamefont{Seidl}},\ }%
  \bibfield{title}{%
  \enquote{\bibinfo {title} {{Simple model for (3+2) neutrino oscillations}},}\
  }%
  \bibfield{journal}{%
  \Doi{10.1016/j.physletb.2004.03.086}{\bibinfo {journal} {Phys. Lett. B}}\ }%
  \textbf{\bibinfo {volume} {591}},\ \bibinfo {pages} {127--136} (\bibinfo
  {year} {2004}),\
  \Eprint{http://arxiv.org/abs/hep-ph/0312285}{arXiv:hep-ph/0312285}%
  \bibAnnoteFile{NoStop}{Babu:2003is}%
\bibitem{Batra:2005rh}%
  \BibitemOpen
  \bibfield{author}{%
  \bibinfo {author} {\bibfnamefont{Puneet}\ \bibnamefont{Batra}}, \bibinfo
  {author} {\bibfnamefont{Bogdan~A.}\ \bibnamefont{Dobrescu}},\ and\ \bibinfo
  {author} {\bibfnamefont{David}\ \bibnamefont{Spivak}},\ }%
  \bibfield{title}{%
  \enquote{\bibinfo {title} {{Anomaly-free sets of fermions}},}\ }%
  \bibfield{journal}{%
  \Doi{10.1063/1.2222081}{\bibinfo {journal} {J. Math. Phys.}}\ }%
  \textbf{\bibinfo {volume} {47}},\ \bibinfo {pages} {082301} (\bibinfo {year}
  {2006}),\ \Eprint{http://arxiv.org/abs/hep-ph/0510181}{arXiv:hep-ph/0510181}%
  \bibAnnoteFile{NoStop}{Batra:2005rh}%
\bibitem{deGouvea:2015pea}%
  \BibitemOpen
  \bibfield{author}{%
  \bibinfo {author} {\bibfnamefont{Andr\'e}\ \bibnamefont{de~Gouv\^ea}}\ and\
  \bibinfo {author} {\bibfnamefont{Daniel}\ \bibnamefont{Hern\'andez}},\ }%
  \bibfield{title}{%
  \enquote{\bibinfo {title} {{New Chiral Fermions, a New Gauge Interaction,
  Dirac Neutrinos, and Dark Matter}},}\ }%
  \bibfield{journal}{%
  \Doi{10.1007/JHEP10(2015)046}{\bibinfo {journal} {JHEP}}\ }%
  \textbf{\bibinfo {volume} {10}},\ \bibinfo {pages} {046} (\bibinfo {year}
  {2015}),\ \Eprint{http://arxiv.org/abs/1507.00916}{arXiv:1507.00916
  [hep-ph]}%
  \bibAnnoteFile{NoStop}{deGouvea:2015pea}%
\bibitem{Costa:2020dph}%
  \BibitemOpen
  \bibfield{author}{%
  \bibinfo {author} {\bibfnamefont{Davi~B.}\ \bibnamefont{Costa}}, \bibinfo
  {author} {\bibfnamefont{Bogdan~A.}\ \bibnamefont{Dobrescu}},\ and\ \bibinfo
  {author} {\bibfnamefont{Patrick~J.}\ \bibnamefont{Fox}},\ }%
  \bibfield{title}{%
  \enquote{\bibinfo {title} {{Chiral Abelian gauge theories with few
  fermions}},}\ }%
  \bibfield{journal}{%
  \Doi{10.1103/PhysRevD.101.095032}{\bibinfo {journal} {Phys. Rev. D}}\ }%
  \textbf{\bibinfo {volume} {101}},\ \bibinfo {pages} {095032} (\bibinfo {year}
  {2020}),\ \Eprint{http://arxiv.org/abs/2001.11991}{arXiv:2001.11991
  [hep-ph]}%
  \bibAnnoteFile{NoStop}{Costa:2020dph}%
\bibitem{delaVega:2023dmw}%
  \BibitemOpen
  \bibfield{author}{%
  \bibinfo {author} {\bibfnamefont{Leon M.~G.}\ \bibnamefont{de~la Vega}},
  \bibinfo {author} {\bibfnamefont{R.}~\bibnamefont{Ferro-Hernandez}}, \bibinfo
  {author} {\bibfnamefont{A.}~\bibnamefont{Garc\'\i{}a-Viltres}}, \bibinfo
  {author} {\bibfnamefont{Eduardo}\ \bibnamefont{Peinado}},\ and\ \bibinfo
  {author} {\bibfnamefont{E.}~\bibnamefont{V\'azquez-J\'auregui}},\ }%
  \bibfield{title}{%
  \enquote{\bibinfo {title} {{Closing the dark photon window to thermal dark
  matter}},}\ }%
   (\bibinfo {month} {11}\ \bibinfo {year} {2023}),\
  \Eprint{http://arxiv.org/abs/2311.17987}{arXiv:2311.17987 [hep-ph]}%
  \bibAnnoteFile{NoStop}{delaVega:2023dmw}%
\bibitem{Ma:2021szi}%
  \BibitemOpen
  \bibfield{author}{%
  \bibinfo {author} {\bibfnamefont{Ernest}\ \bibnamefont{Ma}},\ }%
  \bibfield{title}{%
  \enquote{\bibinfo {title} {{Linkage of Dirac Neutrinos to Dark U(1) Gauge
  Symmetry}},}\ }%
  \bibfield{journal}{%
  \Doi{10.1016/j.physletb.2021.136290}{\bibinfo {journal} {Phys. Lett. B}}\ }%
  \textbf{\bibinfo {volume} {817}},\ \bibinfo {pages} {136290} (\bibinfo {year}
  {2021}),\ \Eprint{http://arxiv.org/abs/2101.12138}{arXiv:2101.12138
  [hep-ph]}%
  \bibAnnoteFile{NoStop}{Ma:2021szi}%
\bibitem{Mention:2011rk}%
  \BibitemOpen
  \bibfield{author}{%
  \bibinfo {author} {\bibfnamefont{G.}~\bibnamefont{Mention}}, \bibinfo
  {author} {\bibfnamefont{M.}~\bibnamefont{Fechner}}, \bibinfo {author}
  {\bibfnamefont{Th.}\ \bibnamefont{Lasserre}}, \bibinfo {author}
  {\bibfnamefont{Th.~A.}\ \bibnamefont{Mueller}}, \bibinfo {author}
  {\bibfnamefont{D.}~\bibnamefont{Lhuillier}}, \bibinfo {author}
  {\bibfnamefont{M.}~\bibnamefont{Cribier}},\ and\ \bibinfo {author}
  {\bibfnamefont{A.}~\bibnamefont{Letourneau}},\ }%
  \bibfield{title}{%
  \enquote{\bibinfo {title} {{The Reactor Antineutrino Anomaly}},}\ }%
  \bibfield{journal}{%
  \Doi{10.1103/PhysRevD.83.073006}{\bibinfo {journal} {Phys. Rev. D}}\ }%
  \textbf{\bibinfo {volume} {83}},\ \bibinfo {pages} {073006} (\bibinfo {year}
  {2011}),\ \Eprint{http://arxiv.org/abs/1101.2755}{arXiv:1101.2755 [hep-ex]}%
  \bibAnnoteFile{NoStop}{Mention:2011rk}%
\bibitem{Davoudiasl:2005ks}%
  \BibitemOpen
  \bibfield{author}{%
  \bibinfo {author} {\bibfnamefont{Hooman}\ \bibnamefont{Davoudiasl}}, \bibinfo
  {author} {\bibfnamefont{Ryuichiro}\ \bibnamefont{Kitano}}, \bibinfo {author}
  {\bibfnamefont{Graham~D.}\ \bibnamefont{Kribs}},\ and\ \bibinfo {author}
  {\bibfnamefont{Hitoshi}\ \bibnamefont{Murayama}},\ }%
  \bibfield{title}{%
  \enquote{\bibinfo {title} {{Models of neutrino mass with a low cutoff
  scale}},}\ }%
  \bibfield{journal}{%
  \Doi{10.1103/PhysRevD.71.113004}{\bibinfo {journal} {Phys. Rev. D}}\ }%
  \textbf{\bibinfo {volume} {71}},\ \bibinfo {pages} {113004} (\bibinfo {year}
  {2005}),\ \Eprint{http://arxiv.org/abs/hep-ph/0502176}{arXiv:hep-ph/0502176}%
  \bibAnnoteFile{NoStop}{Davoudiasl:2005ks}%
\bibitem{Heeck:2012bz}%
  \BibitemOpen
  \bibfield{author}{%
  \bibinfo {author} {\bibfnamefont{Julian}\ \bibnamefont{Heeck}}\ and\ \bibinfo
  {author} {\bibfnamefont{He}~\bibnamefont{Zhang}},\ }%
  \bibfield{title}{%
  \enquote{\bibinfo {title} {{Exotic Charges, Multicomponent Dark Matter and
  Light Sterile Neutrinos}},}\ }%
  \bibfield{journal}{%
  \Doi{10.1007/JHEP05(2013)164}{\bibinfo {journal} {JHEP}}\ }%
  \textbf{\bibinfo {volume} {05}},\ \bibinfo {pages} {164} (\bibinfo {year}
  {2013}),\ \Eprint{http://arxiv.org/abs/1211.0538}{arXiv:1211.0538 [hep-ph]}%
  \bibAnnoteFile{NoStop}{Heeck:2012bz}%
\bibitem{Batell:2010bp}%
  \BibitemOpen
  \bibfield{author}{%
  \bibinfo {author} {\bibfnamefont{Brian}\ \bibnamefont{Batell}},\ }%
  \bibfield{title}{%
  \enquote{\bibinfo {title} {{Dark Discrete Gauge Symmetries}},}\ }%
  \bibfield{journal}{%
  \Doi{10.1103/PhysRevD.83.035006}{\bibinfo {journal} {Phys. Rev. D}}\ }%
  \textbf{\bibinfo {volume} {83}},\ \bibinfo {pages} {035006} (\bibinfo {year}
  {2011}),\ \Eprint{http://arxiv.org/abs/1007.0045}{arXiv:1007.0045 [hep-ph]}%
  \bibAnnoteFile{NoStop}{Batell:2010bp}%
\bibitem{Calle:2019mxn}%
  \BibitemOpen
  \bibfield{author}{%
  \bibinfo {author} {\bibfnamefont{Julian}\ \bibnamefont{Calle}}, \bibinfo
  {author} {\bibfnamefont{Diego}\ \bibnamefont{Restrepo}},\ and\ \bibinfo
  {author} {\bibfnamefont{\'Oscar}\ \bibnamefont{Zapata}},\ }%
  \bibfield{title}{%
  \enquote{\bibinfo {title} {{Dirac neutrino mass generation from a Majorana
  messenger}},}\ }%
  \bibfield{journal}{%
  \Doi{10.1103/PhysRevD.101.035004}{\bibinfo {journal} {Phys. Rev. D}}\ }%
  \textbf{\bibinfo {volume} {101}},\ \bibinfo {pages} {035004} (\bibinfo {year}
  {2020}),\ \Eprint{http://arxiv.org/abs/1909.09574}{arXiv:1909.09574
  [hep-ph]}%
  \bibAnnoteFile{NoStop}{Calle:2019mxn}%
\bibitem{deSalas:2017kay}%
  \BibitemOpen
  \bibfield{author}{%
  \bibinfo {author} {\bibfnamefont{P.~F.}\ \bibnamefont{de~Salas}}, \bibinfo
  {author} {\bibfnamefont{D.~V.}\ \bibnamefont{Forero}}, \bibinfo {author}
  {\bibfnamefont{C.~A.}\ \bibnamefont{Ternes}}, \bibinfo {author}
  {\bibfnamefont{M.}~\bibnamefont{Tortola}},\ and\ \bibinfo {author}
  {\bibfnamefont{J.~W.~F.}\ \bibnamefont{Valle}},\ }%
  \bibfield{title}{%
  \enquote{\bibinfo {title} {{Status of neutrino oscillations 2018: 3$\sigma$
  hint for normal mass ordering and improved CP sensitivity}},}\ }%
  \bibfield{journal}{%
  \Doi{10.1016/j.physletb.2018.06.019}{\bibinfo {journal} {Phys. Lett. B}}\ }%
  \textbf{\bibinfo {volume} {782}},\ \bibinfo {pages} {633--640} (\bibinfo
  {year} {2018}),\ \Eprint{http://arxiv.org/abs/1708.01186}{arXiv:1708.01186
  [hep-ph]}%
  \bibAnnoteFile{NoStop}{deSalas:2017kay}%
\bibitem{Bauer:2018onh}%
  \BibitemOpen
  \bibfield{author}{%
  \bibinfo {author} {\bibfnamefont{Martin}\ \bibnamefont{Bauer}}, \bibinfo
  {author} {\bibfnamefont{Patrick}\ \bibnamefont{Foldenauer}},\ and\ \bibinfo
  {author} {\bibfnamefont{Joerg}\ \bibnamefont{Jaeckel}},\ }%
  \bibfield{title}{%
  \enquote{\bibinfo {title} {{Hunting All the Hidden Photons}},}\ }%
  \bibfield{journal}{%
  \Doi{10.1007/JHEP07(2018)094}{\bibinfo {journal} {JHEP}}\ }%
  \textbf{\bibinfo {volume} {07}},\ \bibinfo {pages} {094} (\bibinfo {year}
  {2018}),\ \Eprint{http://arxiv.org/abs/1803.05466}{arXiv:1803.05466
  [hep-ph]}%
  \bibAnnoteFile{NoStop}{Bauer:2018onh}%
\bibitem{Belanger:2014vza}%
  \BibitemOpen
  \bibfield{author}{%
  \bibinfo {author} {\bibfnamefont{G.}~\bibnamefont{B\'elanger}}, \bibinfo
  {author} {\bibfnamefont{F.}~\bibnamefont{Boudjema}}, \bibinfo {author}
  {\bibfnamefont{A.}~\bibnamefont{Pukhov}},\ and\ \bibinfo {author}
  {\bibfnamefont{A.}~\bibnamefont{Semenov}},\ }%
  \bibfield{title}{%
  \enquote{\bibinfo {title} {{micrOMEGAs4.1: two dark matter candidates}},}\ }%
  \bibfield{journal}{%
  \Doi{10.1016/j.cpc.2015.03.003}{\bibinfo {journal} {Comput. Phys. Commun.}}\
  }%
  \textbf{\bibinfo {volume} {192}},\ \bibinfo {pages} {322--329} (\bibinfo
  {year} {2015}),\ \Eprint{http://arxiv.org/abs/1407.6129}{arXiv:1407.6129
  [hep-ph]}%
  \bibAnnoteFile{NoStop}{Belanger:2014vza}%
\bibitem{Alguero:2023zol}%
  \BibitemOpen
  \bibfield{author}{%
  \bibinfo {author} {\bibfnamefont{G.}~\bibnamefont{Alguero}}, \bibinfo
  {author} {\bibfnamefont{G.}~\bibnamefont{Belanger}}, \bibinfo {author}
  {\bibfnamefont{F.}~\bibnamefont{Boudjema}}, \bibinfo {author}
  {\bibfnamefont{S.}~\bibnamefont{Chakraborti}}, \bibinfo {author}
  {\bibfnamefont{A.}~\bibnamefont{Goudelis}}, \bibinfo {author}
  {\bibfnamefont{S.}~\bibnamefont{Kraml}}, \bibinfo {author}
  {\bibfnamefont{A.}~\bibnamefont{Mjallal}},\ and\ \bibinfo {author}
  {\bibfnamefont{A.}~\bibnamefont{Pukhov}},\ }%
  \bibfield{title}{%
  \enquote{\bibinfo {title} {{micrOMEGAs 6.0: N-component dark matter}},}\ }%
  \bibfield{journal}{%
  \Doi{10.1016/j.cpc.2024.109133}{\bibinfo {journal} {Comput. Phys. Commun.}}\
  }%
  \textbf{\bibinfo {volume} {299}},\ \bibinfo {pages} {109133} (\bibinfo {year}
  {2024}),\ \Eprint{http://arxiv.org/abs/2312.14894}{arXiv:2312.14894
  [hep-ph]}%
  \bibAnnoteFile{NoStop}{Alguero:2023zol}%
\bibitem{Planck:2018vyg}%
  \BibitemOpen
  \bibfield{author}{%
  \bibinfo {author} {\bibfnamefont{N.}~\bibnamefont{Aghanim}} \emph{et~al.}
  (\bibinfo {collaboration} {Planck}),\ }%
  \bibfield{title}{%
  \enquote{\bibinfo {title} {{Planck 2018 results. VI. Cosmological
  parameters}},}\ }%
  \bibfield{journal}{%
  \Doi{10.1051/0004-6361/201833910}{\bibinfo {journal} {Astron. Astrophys.}}\
  }%
  \textbf{\bibinfo {volume} {641}},\ \bibinfo {pages} {A6} (\bibinfo {year}
  {2020}),\ \bibinfo {note} {[Erratum: Astron.Astrophys. 652, C4 (2021)]},\
  \Eprint{http://arxiv.org/abs/1807.06209}{arXiv:1807.06209 [astro-ph.CO]}%
  \bibAnnoteFile{NoStop}{Planck:2018vyg}%
\bibitem{Hannestad:1995rs}%
  \BibitemOpen
  \bibfield{author}{%
  \bibinfo {author} {\bibfnamefont{Steen}\ \bibnamefont{Hannestad}}\ and\
  \bibinfo {author} {\bibfnamefont{Jes}\ \bibnamefont{Madsen}},\ }%
  \bibfield{title}{%
  \enquote{\bibinfo {title} {{Neutrino decoupling in the early universe}},}\ }%
  \bibfield{journal}{%
  \Doi{10.1103/PhysRevD.52.1764}{\bibinfo {journal} {Phys. Rev. D}}\ }%
  \textbf{\bibinfo {volume} {52}},\ \bibinfo {pages} {1764--1769} (\bibinfo
  {year} {1995}),\
  \Eprint{http://arxiv.org/abs/astro-ph/9506015}{arXiv:astro-ph/9506015}%
  \bibAnnoteFile{NoStop}{Hannestad:1995rs}%
\bibitem{Kawasaki:2000en}%
  \BibitemOpen
  \bibfield{author}{%
  \bibinfo {author} {\bibfnamefont{M.}~\bibnamefont{Kawasaki}}, \bibinfo
  {author} {\bibfnamefont{Kazunori}\ \bibnamefont{Kohri}},\ and\ \bibinfo
  {author} {\bibfnamefont{Naoshi}\ \bibnamefont{Sugiyama}},\ }%
  \bibfield{title}{%
  \enquote{\bibinfo {title} {{MeV scale reheating temperature and
  thermalization of neutrino background}},}\ }%
  \bibfield{journal}{%
  \Doi{10.1103/PhysRevD.62.023506}{\bibinfo {journal} {Phys. Rev. D}}\ }%
  \textbf{\bibinfo {volume} {62}},\ \bibinfo {pages} {023506} (\bibinfo {year}
  {2000}),\
  \Eprint{http://arxiv.org/abs/astro-ph/0002127}{arXiv:astro-ph/0002127}%
  \bibAnnoteFile{NoStop}{Kawasaki:2000en}%
\bibitem{Coy:2024itg}%
  \BibitemOpen
  \bibfield{author}{%
  \bibinfo {author} {\bibfnamefont{Rupert}\ \bibnamefont{Coy}}, \bibinfo
  {author} {\bibfnamefont{Jean}\ \bibnamefont{Kimus}},\ and\ \bibinfo {author}
  {\bibfnamefont{Michel H.~G.}\ \bibnamefont{Tytgat}},\ }%
  \bibfield{title}{%
  \enquote{\bibinfo {title} {{Light from darkness: history of a hot dark
  sector}},}\ }%
   (\bibinfo {month} {5}\ \bibinfo {year} {2024}),\
  \Eprint{http://arxiv.org/abs/2405.10792}{arXiv:2405.10792 [hep-ph]}%
  \bibAnnoteFile{NoStop}{Coy:2024itg}%
\bibitem{Chen:2015dka}%
  \BibitemOpen
  \bibfield{author}{%
  \bibinfo {author} {\bibfnamefont{Mu-Chun}\ \bibnamefont{Chen}}, \bibinfo
  {author} {\bibfnamefont{Michael}\ \bibnamefont{Ratz}},\ and\ \bibinfo
  {author} {\bibfnamefont{Andreas}\ \bibnamefont{Trautner}},\ }%
  \bibfield{title}{%
  \enquote{\bibinfo {title} {{Nonthermal cosmic neutrino background}},}\ }%
  \bibfield{journal}{%
  \Doi{10.1103/PhysRevD.92.123006}{\bibinfo {journal} {Phys. Rev. D}}\ }%
  \textbf{\bibinfo {volume} {92}},\ \bibinfo {pages} {123006} (\bibinfo {year}
  {2015}),\ \Eprint{http://arxiv.org/abs/1509.00481}{arXiv:1509.00481
  [hep-ph]}%
  \bibAnnoteFile{NoStop}{Chen:2015dka}%
\bibitem{Kolb:1990vq}%
  \BibitemOpen
  \bibfield{author}{%
  \bibinfo {author} {\bibfnamefont{Edward~W.}\ \bibnamefont{Kolb}}\ and\
  \bibinfo {author} {\bibfnamefont{Michael~S.}\ \bibnamefont{Turner}},\ }%
  \bibfield{title}{%
  \enquote{\bibinfo {title} {{The Early Universe}},}\ }%
  \bibfield{journal}{%
  \bibinfo {journal} {Front. Phys.}\ }%
  \textbf{\bibinfo {volume} {69}},\ \bibinfo {pages} {1--547} (\bibinfo {year}
  {1990})%
  \bibAnnoteFile{NoStop}{Kolb:1990vq}%
\bibitem{Srednicki:1988ce}%
  \BibitemOpen
  \bibfield{author}{%
  \bibinfo {author} {\bibfnamefont{Mark}\ \bibnamefont{Srednicki}}, \bibinfo
  {author} {\bibfnamefont{Richard}\ \bibnamefont{Watkins}},\ and\ \bibinfo
  {author} {\bibfnamefont{Keith~A.}\ \bibnamefont{Olive}},\ }%
  \bibfield{title}{%
  \enquote{\bibinfo {title} {{Calculations of Relic Densities in the Early
  Universe}},}\ }%
  \bibfield{journal}{%
  \Doi{10.1016/0550-3213(88)90099-5}{\bibinfo {journal} {Nucl. Phys.}}\ }%
  \textbf{\bibinfo {volume} {B310}},\ \bibinfo {pages} {693} (\bibinfo {year}
  {1988}),\ \bibinfo {note} {[,247(1988)]}%
  \bibAnnoteFile{NoStop}{Srednicki:1988ce}%
\bibitem{Hahn:2000kx}%
  \BibitemOpen
  \bibfield{author}{%
  \bibinfo {author} {\bibfnamefont{Thomas}\ \bibnamefont{Hahn}},\ }%
  \bibfield{title}{%
  \enquote{\bibinfo {title} {{Generating Feynman diagrams and amplitudes with
  FeynArts 3}},}\ }%
  \bibfield{journal}{%
  \Doi{10.1016/S0010-4655(01)00290-9}{\bibinfo {journal} {Comput. Phys.
  Commun.}}\ }%
  \textbf{\bibinfo {volume} {140}},\ \bibinfo {pages} {418--431} (\bibinfo
  {year} {2001}),\
  \Eprint{http://arxiv.org/abs/hep-ph/0012260}{arXiv:hep-ph/0012260}%
  \bibAnnoteFile{NoStop}{Hahn:2000kx}%
\bibitem{Mertig:1990an}%
  \BibitemOpen
  \bibfield{author}{%
  \bibinfo {author} {\bibfnamefont{R.}~\bibnamefont{Mertig}}, \bibinfo {author}
  {\bibfnamefont{M.}~\bibnamefont{Bohm}},\ and\ \bibinfo {author}
  {\bibfnamefont{Ansgar}\ \bibnamefont{Denner}},\ }%
  \bibfield{title}{%
  \enquote{\bibinfo {title} {{FEYN CALC: Computer algebraic calculation of
  Feynman amplitudes}},}\ }%
  \bibfield{journal}{%
  \Doi{10.1016/0010-4655(91)90130-D}{\bibinfo {journal} {Comput. Phys.
  Commun.}}\ }%
  \textbf{\bibinfo {volume} {64}},\ \bibinfo {pages} {345--359} (\bibinfo
  {year} {1991})%
  \bibAnnoteFile{NoStop}{Mertig:1990an}%
\bibitem{Shtabovenko:2016sxi}%
  \BibitemOpen
  \bibfield{author}{%
  \bibinfo {author} {\bibfnamefont{Vladyslav}\ \bibnamefont{Shtabovenko}},
  \bibinfo {author} {\bibfnamefont{Rolf}\ \bibnamefont{Mertig}},\ and\ \bibinfo
  {author} {\bibfnamefont{Frederik}\ \bibnamefont{Orellana}},\ }%
  \bibfield{title}{%
  \enquote{\bibinfo {title} {{New Developments in FeynCalc 9.0}},}\ }%
  \bibfield{journal}{%
  \Doi{10.1016/j.cpc.2016.06.008}{\bibinfo {journal} {Comput. Phys. Commun.}}\
  }%
  \textbf{\bibinfo {volume} {207}},\ \bibinfo {pages} {432--444} (\bibinfo
  {year} {2016}),\ \Eprint{http://arxiv.org/abs/1601.01167}{arXiv:1601.01167
  [hep-ph]}%
  \bibAnnoteFile{NoStop}{Shtabovenko:2016sxi}%
\bibitem{Shtabovenko:2020gxv}%
  \BibitemOpen
  \bibfield{author}{%
  \bibinfo {author} {\bibfnamefont{Vladyslav}\ \bibnamefont{Shtabovenko}},
  \bibinfo {author} {\bibfnamefont{Rolf}\ \bibnamefont{Mertig}},\ and\ \bibinfo
  {author} {\bibfnamefont{Frederik}\ \bibnamefont{Orellana}},\ }%
  \bibfield{title}{%
  \enquote{\bibinfo {title} {{FeynCalc 9.3: New features and improvements}},}\
  }%
  \bibfield{journal}{%
  \Doi{10.1016/j.cpc.2020.107478}{\bibinfo {journal} {Comput. Phys. Commun.}}\
  }%
  \textbf{\bibinfo {volume} {256}},\ \bibinfo {pages} {107478} (\bibinfo {year}
  {2020}),\ \Eprint{http://arxiv.org/abs/2001.04407}{arXiv:2001.04407
  [hep-ph]}%
  \bibAnnoteFile{NoStop}{Shtabovenko:2020gxv}%
\bibitem{Ma:2022uhi}%
  \BibitemOpen
  \bibfield{author}{%
  \bibinfo {author} {\bibfnamefont{Ernest}\ \bibnamefont{Ma}},\ }%
  \bibfield{title}{%
  \enquote{\bibinfo {title} {{Connecting dark gauge symmetry to the standard
  model}},}\ }%
  \bibfield{journal}{%
  \Doi{10.1016/j.physletb.2022.137282}{\bibinfo {journal} {Phys. Lett. B}}\ }%
  \textbf{\bibinfo {volume} {833}},\ \bibinfo {pages} {137282} (\bibinfo {year}
  {2022}),\ \Eprint{http://arxiv.org/abs/2203.12034}{2203.12034 [hep-ph]}%
  \bibAnnoteFile{NoStop}{Ma:2022uhi}%
\bibitem{Alves:2015pea}%
  \BibitemOpen
  \bibfield{author}{%
  \bibinfo {author} {\bibfnamefont{Alexandre}\ \bibnamefont{Alves}}, \bibinfo
  {author} {\bibfnamefont{Asher}\ \bibnamefont{Berlin}}, \bibinfo {author}
  {\bibfnamefont{Stefano}\ \bibnamefont{Profumo}},\ and\ \bibinfo {author}
  {\bibfnamefont{Farinaldo~S.}\ \bibnamefont{Queiroz}},\ }%
  \bibfield{title}{%
  \enquote{\bibinfo {title} {{Dark Matter Complementarity and the Z$^\prime$
  Portal}},}\ }%
  \bibfield{journal}{%
  \Doi{10.1103/PhysRevD.92.083004}{\bibinfo {journal} {Phys. Rev. D}}\ }%
  \textbf{\bibinfo {volume} {92}},\ \bibinfo {pages} {083004} (\bibinfo {year}
  {2015}),\ \Eprint{http://arxiv.org/abs/1501.03490}{arXiv:1501.03490
  [hep-ph]}%
  \bibAnnoteFile{NoStop}{Alves:2015pea}%
\bibitem{Alves:2015mua}%
  \BibitemOpen
  \bibfield{author}{%
  \bibinfo {author} {\bibfnamefont{Alexandre}\ \bibnamefont{Alves}}, \bibinfo
  {author} {\bibfnamefont{Asher}\ \bibnamefont{Berlin}}, \bibinfo {author}
  {\bibfnamefont{Stefano}\ \bibnamefont{Profumo}},\ and\ \bibinfo {author}
  {\bibfnamefont{Farinaldo~S.}\ \bibnamefont{Queiroz}},\ }%
  \bibfield{title}{%
  \enquote{\bibinfo {title} {{Dirac-fermionic dark matter in U(1)$_{X}$
  models}},}\ }%
  \bibfield{journal}{%
  \Doi{10.1007/JHEP10(2015)076}{\bibinfo {journal} {JHEP}}\ }%
  \textbf{\bibinfo {volume} {10}},\ \bibinfo {pages} {076} (\bibinfo {year}
  {2015}),\ \Eprint{http://arxiv.org/abs/1506.06767}{arXiv:1506.06767
  [hep-ph]}%
  \bibAnnoteFile{NoStop}{Alves:2015mua}%
\bibitem{Staub:2008uz}%
  \BibitemOpen
  \bibfield{author}{%
  \bibinfo {author} {\bibfnamefont{F.}~\bibnamefont{Staub}},\ }%
  \bibfield{title}{%
  \enquote{\bibinfo {title} {{SARAH}},}\ }%
   (\bibinfo {month} {6}\ \bibinfo {year} {2008}),\
  \Eprint{http://arxiv.org/abs/0806.0538}{arXiv:0806.0538 [hep-ph]}%
  \bibAnnoteFile{NoStop}{Staub:2008uz}%
\bibitem{Staub:2009bi}%
  \BibitemOpen
  \bibfield{author}{%
  \bibinfo {author} {\bibfnamefont{Florian}\ \bibnamefont{Staub}},\ }%
  \bibfield{title}{%
  \enquote{\bibinfo {title} {{From Superpotential to Model Files for FeynArts
  and CalcHep/CompHep}},}\ }%
  \bibfield{journal}{%
  \Doi{10.1016/j.cpc.2010.01.011}{\bibinfo {journal} {Comput. Phys. Commun.}}\
  }%
  \textbf{\bibinfo {volume} {181}},\ \bibinfo {pages} {1077--1086} (\bibinfo
  {year} {2010}),\ \Eprint{http://arxiv.org/abs/0909.2863}{arXiv:0909.2863
  [hep-ph]}%
  \bibAnnoteFile{NoStop}{Staub:2009bi}%
\bibitem{Staub:2010jh}%
  \BibitemOpen
  \bibfield{author}{%
  \bibinfo {author} {\bibfnamefont{Florian}\ \bibnamefont{Staub}},\ }%
  \bibfield{title}{%
  \enquote{\bibinfo {title} {{Automatic Calculation of supersymmetric
  Renormalization Group Equations and Self Energies}},}\ }%
  \bibfield{journal}{%
  \Doi{10.1016/j.cpc.2010.11.030}{\bibinfo {journal} {Comput. Phys. Commun.}}\
  }%
  \textbf{\bibinfo {volume} {182}},\ \bibinfo {pages} {808--833} (\bibinfo
  {year} {2011}),\ \Eprint{http://arxiv.org/abs/1002.0840}{arXiv:1002.0840
  [hep-ph]}%
  \bibAnnoteFile{NoStop}{Staub:2010jh}%
\bibitem{Staub:2012pb}%
  \BibitemOpen
  \bibfield{author}{%
  \bibinfo {author} {\bibfnamefont{Florian}\ \bibnamefont{Staub}},\ }%
  \bibfield{title}{%
  \enquote{\bibinfo {title} {{SARAH 3.2: Dirac Gauginos, UFO output, and
  more}},}\ }%
  \bibfield{journal}{%
  \Doi{10.1016/j.cpc.2013.02.019}{\bibinfo {journal} {Comput. Phys. Commun.}}\
  }%
  \textbf{\bibinfo {volume} {184}},\ \bibinfo {pages} {1792--1809} (\bibinfo
  {year} {2013}),\ \Eprint{http://arxiv.org/abs/1207.0906}{arXiv:1207.0906
  [hep-ph]}%
  \bibAnnoteFile{NoStop}{Staub:2012pb}%
\bibitem{Staub:2013tta}%
  \BibitemOpen
  \bibfield{author}{%
  \bibinfo {author} {\bibfnamefont{Florian}\ \bibnamefont{Staub}},\ }%
  \bibfield{title}{%
  \enquote{\bibinfo {title} {{SARAH 4 : A tool for (not only SUSY) model
  builders}},}\ }%
  \bibfield{journal}{%
  \Doi{10.1016/j.cpc.2014.02.018}{\bibinfo {journal} {Comput. Phys. Commun.}}\
  }%
  \textbf{\bibinfo {volume} {185}},\ \bibinfo {pages} {1773--1790} (\bibinfo
  {year} {2014}),\ \Eprint{http://arxiv.org/abs/1309.7223}{arXiv:1309.7223
  [hep-ph]}%
  \bibAnnoteFile{NoStop}{Staub:2013tta}%
\bibitem{Porod:2003um}%
  \BibitemOpen
  \bibfield{author}{%
  \bibinfo {author} {\bibfnamefont{Werner}\ \bibnamefont{Porod}},\ }%
  \bibfield{title}{%
  \enquote{\bibinfo {title} {{SPheno, a program for calculating supersymmetric
  spectra, SUSY particle decays and SUSY particle production at e+ e-
  colliders}},}\ }%
  \bibfield{journal}{%
  \Doi{10.1016/S0010-4655(03)00222-4}{\bibinfo {journal} {Comput. Phys.
  Commun.}}\ }%
  \textbf{\bibinfo {volume} {153}},\ \bibinfo {pages} {275--315} (\bibinfo
  {year} {2003}),\
  \Eprint{http://arxiv.org/abs/hep-ph/0301101}{arXiv:hep-ph/0301101}%
  \bibAnnoteFile{NoStop}{Porod:2003um}%
\bibitem{Porod:2011nf}%
  \BibitemOpen
  \bibfield{author}{%
  \bibinfo {author} {\bibfnamefont{W.}~\bibnamefont{Porod}}\ and\ \bibinfo
  {author} {\bibfnamefont{F.}~\bibnamefont{Staub}},\ }%
  \bibfield{title}{%
  \enquote{\bibinfo {title} {{SPheno 3.1: Extensions including flavour,
  CP-phases and models beyond the MSSM}},}\ }%
  \bibfield{journal}{%
  \Doi{10.1016/j.cpc.2012.05.021}{\bibinfo {journal} {Comput. Phys. Commun.}}\
  }%
  \textbf{\bibinfo {volume} {183}},\ \bibinfo {pages} {2458--2469} (\bibinfo
  {year} {2012}),\ \Eprint{http://arxiv.org/abs/1104.1573}{arXiv:1104.1573
  [hep-ph]}%
  \bibAnnoteFile{NoStop}{Porod:2011nf}%
\bibitem{Falkowski:2015iwa}%
  \BibitemOpen
  \bibfield{author}{%
  \bibinfo {author} {\bibfnamefont{Adam}\ \bibnamefont{Falkowski}}, \bibinfo
  {author} {\bibfnamefont{Christian}\ \bibnamefont{Gross}},\ and\ \bibinfo
  {author} {\bibfnamefont{Oleg}\ \bibnamefont{Lebedev}},\ }%
  \bibfield{title}{%
  \enquote{\bibinfo {title} {{A second Higgs from the Higgs portal}},}\ }%
  \bibfield{journal}{%
  \Doi{10.1007/JHEP05(2015)057}{\bibinfo {journal} {JHEP}}\ }%
  \textbf{\bibinfo {volume} {05}},\ \bibinfo {pages} {057} (\bibinfo {year}
  {2015}),\ \Eprint{http://arxiv.org/abs/1502.01361}{arXiv:1502.01361
  [hep-ph]}%
  \bibAnnoteFile{NoStop}{Falkowski:2015iwa}%
\bibitem{Arcadi:2019lka}%
  \BibitemOpen
  \bibfield{author}{%
  \bibinfo {author} {\bibfnamefont{Giorgio}\ \bibnamefont{Arcadi}}, \bibinfo
  {author} {\bibfnamefont{Abdelhak}\ \bibnamefont{Djouadi}},\ and\ \bibinfo
  {author} {\bibfnamefont{Martti}\ \bibnamefont{Raidal}},\ }%
  \bibfield{title}{%
  \enquote{\bibinfo {title} {{Dark Matter through the Higgs portal}},}\ }%
  \bibfield{journal}{%
  \Doi{10.1016/j.physrep.2019.11.003}{\bibinfo {journal} {Phys. Rept.}}\ }%
  \textbf{\bibinfo {volume} {842}},\ \bibinfo {pages} {1--180} (\bibinfo {year}
  {2020}),\ \Eprint{http://arxiv.org/abs/1903.03616}{arXiv:1903.03616
  [hep-ph]}%
  \bibAnnoteFile{NoStop}{Arcadi:2019lka}%
\bibitem{Ferber:2023iso}%
  \BibitemOpen
  \bibfield{author}{%
  \bibinfo {author} {\bibfnamefont{Torben}\ \bibnamefont{Ferber}}, \bibinfo
  {author} {\bibfnamefont{Alexander}\ \bibnamefont{Grohsjean}},\ and\ \bibinfo
  {author} {\bibfnamefont{Felix}\ \bibnamefont{Kahlhoefer}},\ }%
  \bibfield{title}{%
  \enquote{\bibinfo {title} {{Dark Higgs bosons at colliders}},}\ }%
  \bibfield{journal}{%
  \Doi{10.1016/j.ppnp.2024.104105}{\bibinfo {journal} {Prog. Part. Nucl.
  Phys.}}\ }%
  \textbf{\bibinfo {volume} {136}},\ \bibinfo {pages} {104105} (\bibinfo {year}
  {2024}),\ \Eprint{http://arxiv.org/abs/2305.16169}{arXiv:2305.16169
  [hep-ph]}%
  \bibAnnoteFile{NoStop}{Ferber:2023iso}%
\bibitem{CMS:2022dwd}%
  \BibitemOpen
  \bibfield{author}{%
  \bibinfo {author} {\bibfnamefont{Armen}\ \bibnamefont{Tumasyan}}
  \emph{et~al.} (\bibinfo {collaboration} {CMS}),\ }%
  \bibfield{title}{%
  \enquote{\bibinfo {title} {{A portrait of the Higgs boson by the CMS
  experiment ten years after the discovery.}}.}\ }%
  \bibfield{journal}{%
  \Doi{10.1038/s41586-022-04892-x}{\bibinfo {journal} {Nature}}\ }%
  \textbf{\bibinfo {volume} {607}},\ \bibinfo {pages} {60--68} (\bibinfo {year}
  {2022}),\ \bibinfo {note} {[Erratum: Nature 623, (2023)]},\
  \Eprint{http://arxiv.org/abs/2207.00043}{arXiv:2207.00043 [hep-ex]}%
  \bibAnnoteFile{NoStop}{CMS:2022dwd}%
\bibitem{Babu:2024zoe}%
  \BibitemOpen
  \bibfield{author}{%
  \bibinfo {author} {\bibfnamefont{K.~S.}\ \bibnamefont{Babu}}, \bibinfo
  {author} {\bibfnamefont{Shreyashi}\ \bibnamefont{Chakdar}},\ and\ \bibinfo
  {author} {\bibfnamefont{Vishnu~P.}\ \bibnamefont{K}},\ }%
  \bibfield{title}{%
  \enquote{\bibinfo {title} {{Chiral dark matter and radiative neutrino masses
  from gauged U(1) symmetry}},}\ }%
   (\bibinfo {month} {9}\ \bibinfo {year} {2024}),\
  \Eprint{http://arxiv.org/abs/2409.09008}{arXiv:2409.09008 [hep-ph]}%
  \bibAnnoteFile{NoStop}{Babu:2024zoe}%
\bibitem{Maki:1962mu}%
  \BibitemOpen
  \bibfield{author}{%
  \bibinfo {author} {\bibfnamefont{Ziro}\ \bibnamefont{Maki}}, \bibinfo
  {author} {\bibfnamefont{Masami}\ \bibnamefont{Nakagawa}},\ and\ \bibinfo
  {author} {\bibfnamefont{Shoichi}\ \bibnamefont{Sakata}},\ }%
  \bibfield{title}{%
  \enquote{\bibinfo {title} {{Remarks on the unified model of elementary
  particles}},}\ }%
  \bibfield{journal}{%
  \Doi{10.1143/PTP.28.870}{\bibinfo {journal} {Prog. Theor. Phys.}}\ }%
  \textbf{\bibinfo {volume} {28}},\ \bibinfo {pages} {870--880} (\bibinfo
  {year} {1962})%
  \bibAnnoteFile{NoStop}{Maki:1962mu}%
\bibitem{Duerr:2014wra}%
  \BibitemOpen
  \bibfield{author}{%
  \bibinfo {author} {\bibfnamefont{Michael}\ \bibnamefont{Duerr}}\ and\
  \bibinfo {author} {\bibfnamefont{Pavel}\ \bibnamefont{Fileviez~Perez}},\ }%
  \bibfield{title}{%
  \enquote{\bibinfo {title} {{Theory for Baryon Number and Dark Matter at the
  LHC}},}\ }%
  \bibfield{journal}{%
  \Doi{10.1103/PhysRevD.91.095001}{\bibinfo {journal} {Phys. Rev. D}}\ }%
  \textbf{\bibinfo {volume} {91}},\ \bibinfo {pages} {095001} (\bibinfo {year}
  {2015}),\ \Eprint{http://arxiv.org/abs/1409.8165}{arXiv:1409.8165 [hep-ph]}%
  \bibAnnoteFile{NoStop}{Duerr:2014wra}%
\bibitem{Restrepo:2022cpq}%
  \BibitemOpen
  \bibfield{author}{%
  \bibinfo {author} {\bibfnamefont{Diego}\ \bibnamefont{Restrepo}}, \bibinfo
  {author} {\bibfnamefont{Andr\'es}\ \bibnamefont{Rivera}},\ and\ \bibinfo
  {author} {\bibfnamefont{Walter}\ \bibnamefont{Tangarife}},\ }%
  \bibfield{title}{%
  \enquote{\bibinfo {title} {{Dirac dark matter, neutrino masses, and dark
  baryogenesis}},}\ }%
  \bibfield{journal}{%
  \Doi{10.1103/PhysRevD.106.055021}{\bibinfo {journal} {Phys. Rev. D}}\ }%
  \textbf{\bibinfo {volume} {106}},\ \bibinfo {pages} {055021} (\bibinfo {year}
  {2022}),\ \Eprint{http://arxiv.org/abs/2205.05762}{arXiv:2205.05762
  [hep-ph]}%
  \bibAnnoteFile{NoStop}{Restrepo:2022cpq}%
\bibitem{Yaguna:2024jor}%
  \BibitemOpen
  \bibfield{author}{%
  \bibinfo {author} {\bibfnamefont{Carlos~E.}\ \bibnamefont{Yaguna}}\ and\
  \bibinfo {author} {\bibfnamefont{\'Oscar}\ \bibnamefont{Zapata}},\ }%
  \bibfield{title}{%
  \enquote{\bibinfo {title} {{Singlet Dirac dark matter streamlined}},}\ }%
  \bibfield{journal}{%
  \Doi{10.1088/1475-7516/2024/06/049}{\bibinfo {journal} {JCAP}}\ }%
  \textbf{\bibinfo {volume} {06}},\ \bibinfo {pages} {049} (\bibinfo {year}
  {2024}),\ \Eprint{http://arxiv.org/abs/2401.13101}{arXiv:2401.13101
  [hep-ph]}%
  \bibAnnoteFile{NoStop}{Yaguna:2024jor}%
\bibitem{XENON:2023cxc}%
  \BibitemOpen
  \bibfield{author}{%
  \bibinfo {author} {\bibfnamefont{E.}~\bibnamefont{Aprile}} \emph{et~al.}
  (\bibinfo {collaboration} {XENON}),\ }%
  \bibfield{title}{%
  \enquote{\bibinfo {title} {{First Dark Matter Search with Nuclear Recoils
  from the XENONnT Experiment}},}\ }%
  \bibfield{journal}{%
  \Doi{10.1103/PhysRevLett.131.041003}{\bibinfo {journal} {Phys. Rev. Lett.}}\
  }%
  \textbf{\bibinfo {volume} {131}},\ \bibinfo {pages} {041003} (\bibinfo {year}
  {2023}),\ \Eprint{http://arxiv.org/abs/2303.14729}{arXiv:2303.14729
  [hep-ex]}%
  \bibAnnoteFile{NoStop}{XENON:2023cxc}%
\bibitem{LZ:2022lsv}%
  \BibitemOpen
  \bibfield{author}{%
  \bibinfo {author} {\bibfnamefont{J.}~\bibnamefont{Aalbers}} \emph{et~al.}
  (\bibinfo {collaboration} {LZ}),\ }%
  \bibfield{title}{%
  \enquote{\bibinfo {title} {{First Dark Matter Search Results from the
  LUX-ZEPLIN (LZ) Experiment}},}\ }%
  \bibfield{journal}{%
  \Doi{10.1103/PhysRevLett.131.041002}{\bibinfo {journal} {Phys. Rev. Lett.}}\
  }%
  \textbf{\bibinfo {volume} {131}},\ \bibinfo {pages} {041002} (\bibinfo {year}
  {2023}),\ \Eprint{http://arxiv.org/abs/2207.03764}{arXiv:2207.03764
  [hep-ex]}%
  \bibAnnoteFile{NoStop}{LZ:2022lsv}%
\bibitem{DARWIN:2016hyl}%
  \BibitemOpen
  \bibfield{author}{%
  \bibinfo {author} {\bibfnamefont{J.}~\bibnamefont{Aalbers}} \emph{et~al.}
  (\bibinfo {collaboration} {DARWIN}),\ }%
  \bibfield{title}{%
  \enquote{\bibinfo {title} {{DARWIN: towards the ultimate dark matter
  detector}},}\ }%
  \bibfield{journal}{%
  \Doi{10.1088/1475-7516/2016/11/017}{\bibinfo {journal} {JCAP}}\ }%
  \textbf{\bibinfo {volume} {11}},\ \bibinfo {pages} {017} (\bibinfo {year}
  {2016}),\ \Eprint{http://arxiv.org/abs/1606.07001}{arXiv:1606.07001
  [astro-ph.IM]}%
  \bibAnnoteFile{NoStop}{DARWIN:2016hyl}%
\bibitem{Billard:2013qya}%
  \BibitemOpen
  \bibfield{author}{%
  \bibinfo {author} {\bibfnamefont{J.}~\bibnamefont{Billard}}, \bibinfo
  {author} {\bibfnamefont{L.}~\bibnamefont{Strigari}},\ and\ \bibinfo {author}
  {\bibfnamefont{E.}~\bibnamefont{Figueroa-Feliciano}},\ }%
  \bibfield{title}{%
  \enquote{\bibinfo {title} {{Implication of neutrino backgrounds on the reach
  of next generation dark matter direct detection experiments}},}\ }%
  \bibfield{journal}{%
  \Doi{10.1103/PhysRevD.89.023524}{\bibinfo {journal} {Phys. Rev. D}}\ }%
  \textbf{\bibinfo {volume} {89}},\ \bibinfo {pages} {023524} (\bibinfo {year}
  {2014}),\ \Eprint{http://arxiv.org/abs/1307.5458}{arXiv:1307.5458 [hep-ph]}%
  \bibAnnoteFile{NoStop}{Billard:2013qya}%
\bibitem{Biswas:2022fga}%
  \BibitemOpen
  \bibfield{author}{%
  \bibinfo {author} {\bibfnamefont{Anirban}\ \bibnamefont{Biswas}}, \bibinfo
  {author} {\bibfnamefont{Dilip~Kumar}\ \bibnamefont{Ghosh}},\ and\ \bibinfo
  {author} {\bibfnamefont{Dibyendu}\ \bibnamefont{Nanda}},\ }%
  \bibfield{title}{%
  \enquote{\bibinfo {title} {{Concealing Dirac neutrinos from cosmic microwave
  background}},}\ }%
  \bibfield{journal}{%
  \Doi{10.1088/1475-7516/2022/10/006}{\bibinfo {journal} {JCAP}}\ }%
  \textbf{\bibinfo {volume} {10}},\ \bibinfo {pages} {006} (\bibinfo {year}
  {2022}),\ \Eprint{http://arxiv.org/abs/2206.13710}{arXiv:2206.13710
  [hep-ph]}%
  \bibAnnoteFile{NoStop}{Biswas:2022fga}%
\bibitem{XENON:2019rxp}%
  \BibitemOpen
  \bibfield{author}{%
  \bibinfo {author} {\bibfnamefont{E.}~\bibnamefont{Aprile}} \emph{et~al.}
  (\bibinfo {collaboration} {XENON}),\ }%
  \bibfield{title}{%
  \enquote{\bibinfo {title} {{Constraining the spin-dependent WIMP-nucleon
  cross sections with XENON1T}},}\ }%
  \bibfield{journal}{%
  \Doi{10.1103/PhysRevLett.122.141301}{\bibinfo {journal} {Phys. Rev. Lett.}}\
  }%
  \textbf{\bibinfo {volume} {122}},\ \bibinfo {pages} {141301} (\bibinfo {year}
  {2019}),\ \Eprint{http://arxiv.org/abs/1902.03234}{arXiv:1902.03234
  [astro-ph.CO]}%
  \bibAnnoteFile{NoStop}{XENON:2019rxp}%
\bibitem{Mardon:2009rc}%
  \BibitemOpen
  \bibfield{author}{%
  \bibinfo {author} {\bibfnamefont{Jeremy}\ \bibnamefont{Mardon}}, \bibinfo
  {author} {\bibfnamefont{Yasunori}\ \bibnamefont{Nomura}}, \bibinfo {author}
  {\bibfnamefont{Daniel}\ \bibnamefont{Stolarski}},\ and\ \bibinfo {author}
  {\bibfnamefont{Jesse}\ \bibnamefont{Thaler}},\ }%
  \bibfield{title}{%
  \enquote{\bibinfo {title} {{Dark Matter Signals from Cascade
  Annihilations}},}\ }%
  \bibfield{journal}{%
  \Doi{10.1088/1475-7516/2009/05/016}{\bibinfo {journal} {JCAP}}\ }%
  \textbf{\bibinfo {volume} {05}},\ \bibinfo {pages} {016} (\bibinfo {year}
  {2009}),\ \Eprint{http://arxiv.org/abs/0901.2926}{arXiv:0901.2926 [hep-ph]}%
  \bibAnnoteFile{NoStop}{Mardon:2009rc}%
\bibitem{Siqueira:2021lqj}%
  \BibitemOpen
  \bibfield{author}{%
  \bibinfo {author} {\bibfnamefont{Clarissa}\ \bibnamefont{Siqueira}}, \bibinfo
  {author} {\bibfnamefont{Guilherme~N.}\ \bibnamefont{Fortes}}, \bibinfo
  {author} {\bibfnamefont{Aion}\ \bibnamefont{Viana}},\ and\ \bibinfo {author}
  {\bibfnamefont{Farinaldo~S.}\ \bibnamefont{Queiroz}},\ }%
  \bibfield{title}{%
  \enquote{\bibinfo {title} {{Indirect Searches for Secluded Dark Matter}},}\
  }%
  \bibfield{journal}{%
  \Doi{10.22323/1.395.0577}{\bibinfo {journal} {PoS}}\ }%
  \textbf{\bibinfo {volume} {ICRC2021}},\ \bibinfo {pages} {577} (\bibinfo
  {year} {2021}),\ \Eprint{http://arxiv.org/abs/2107.04053}{arXiv:2107.04053
  [hep-ph]}%
  \bibAnnoteFile{NoStop}{Siqueira:2021lqj}%
\bibitem{Fermi-LAT:2015att}%
  \BibitemOpen
  \bibfield{author}{%
  \bibinfo {author} {\bibfnamefont{M.}~\bibnamefont{Ackermann}} \emph{et~al.}
  (\bibinfo {collaboration} {Fermi-LAT}),\ }%
  \bibfield{title}{%
  \enquote{\bibinfo {title} {{Searching for Dark Matter Annihilation from Milky
  Way Dwarf Spheroidal Galaxies with Six Years of Fermi Large Area Telescope
  Data}},}\ }%
  \bibfield{journal}{%
  \Doi{10.1103/PhysRevLett.115.231301}{\bibinfo {journal} {Phys. Rev. Lett.}}\
  }%
  \textbf{\bibinfo {volume} {115}},\ \bibinfo {pages} {231301} (\bibinfo {year}
  {2015}),\ \Eprint{http://arxiv.org/abs/1503.02641}{arXiv:1503.02641
  [astro-ph.HE]}%
  \bibAnnoteFile{NoStop}{Fermi-LAT:2015att}%
\bibitem{HESS:2016mib}%
  \BibitemOpen
  \bibfield{author}{%
  \bibinfo {author} {\bibfnamefont{H.}~\bibnamefont{Abdallah}} \emph{et~al.}
  (\bibinfo {collaboration} {H.E.S.S.}),\ }%
  \bibfield{title}{%
  \enquote{\bibinfo {title} {{Search for dark matter annihilations towards the
  inner Galactic halo from 10 years of observations with H.E.S.S}},}\ }%
  \bibfield{journal}{%
  \Doi{10.1103/PhysRevLett.117.111301}{\bibinfo {journal} {Phys. Rev. Lett.}}\
  }%
  \textbf{\bibinfo {volume} {117}},\ \bibinfo {pages} {111301} (\bibinfo {year}
  {2016}),\ \Eprint{http://arxiv.org/abs/1607.08142}{arXiv:1607.08142
  [astro-ph.HE]}%
  \bibAnnoteFile{NoStop}{HESS:2016mib}%
\bibitem{CTA:2020qlo}%
  \BibitemOpen
  \bibfield{author}{%
  \bibinfo {author} {\bibfnamefont{A.}~\bibnamefont{Acharyya}} \emph{et~al.}
  (\bibinfo {collaboration} {CTA}),\ }%
  \bibfield{title}{%
  \enquote{\bibinfo {title} {{Sensitivity of the Cherenkov Telescope Array to a
  dark matter signal from the Galactic centre}},}\ }%
  \bibfield{journal}{%
  \Doi{10.1088/1475-7516/2021/01/057}{\bibinfo {journal} {JCAP}}\ }%
  \textbf{\bibinfo {volume} {01}},\ \bibinfo {pages} {057} (\bibinfo {year}
  {2021}),\ \Eprint{http://arxiv.org/abs/2007.16129}{arXiv:2007.16129
  [astro-ph.HE]}%
  \bibAnnoteFile{NoStop}{CTA:2020qlo}%
\bibitem{Schoorlemmer:2019gee}%
  \BibitemOpen
  \bibfield{author}{%
  \bibinfo {author} {\bibfnamefont{H.}~\bibnamefont{Schoorlemmer}} (\bibinfo
  {collaboration} {SWGO}),\ }%
  \bibfield{title}{%
  \enquote{\bibinfo {title} {{A next-generation ground-based wide field-of-view
  gamma-ray observatory in the southern hemisphere}},}\ }%
  \bibfield{journal}{%
  \Doi{10.22323/1.358.0785}{\bibinfo {journal} {PoS}}\ }%
  \textbf{\bibinfo {volume} {ICRC2019}},\ \bibinfo {pages} {785} (\bibinfo
  {year} {2020}),\ \Eprint{http://arxiv.org/abs/1908.08858}{arXiv:1908.08858
  [astro-ph.HE]}%
  \bibAnnoteFile{NoStop}{Schoorlemmer:2019gee}%
\bibitem{Niblaeus:2019gjk}%
  \BibitemOpen
  \bibfield{author}{%
  \bibinfo {author} {\bibfnamefont{Carl}\ \bibnamefont{Niblaeus}}, \bibinfo
  {author} {\bibfnamefont{Ankit}\ \bibnamefont{Beniwal}},\ and\ \bibinfo
  {author} {\bibfnamefont{Joakim}\ \bibnamefont{Edsjo}},\ }%
  \bibfield{title}{%
  \enquote{\bibinfo {title} {{Neutrinos and gamma rays from long-lived mediator
  decays in the Sun}},}\ }%
  \bibfield{journal}{%
  \Doi{10.1088/1475-7516/2019/11/011}{\bibinfo {journal} {JCAP}}\ }%
  \textbf{\bibinfo {volume} {11}},\ \bibinfo {pages} {011} (\bibinfo {year}
  {2019}),\ \Eprint{http://arxiv.org/abs/1903.11363}{arXiv:1903.11363
  [astro-ph.HE]}%
  \bibAnnoteFile{NoStop}{Niblaeus:2019gjk}%
\bibitem{Bell:2021pyy}%
  \BibitemOpen
  \bibfield{author}{%
  \bibinfo {author} {\bibfnamefont{Nicole~F.}\ \bibnamefont{Bell}}, \bibinfo
  {author} {\bibfnamefont{James~B.}\ \bibnamefont{Dent}},\ and\ \bibinfo
  {author} {\bibfnamefont{Isaac~W.}\ \bibnamefont{Sanderson}},\ }%
  \bibfield{title}{%
  \enquote{\bibinfo {title} {{Solar gamma ray constraints on dark matter
  annihilation to secluded mediators}},}\ }%
  \bibfield{journal}{%
  \Doi{10.1103/PhysRevD.104.023024}{\bibinfo {journal} {Phys. Rev. D}}\ }%
  \textbf{\bibinfo {volume} {104}},\ \bibinfo {pages} {023024} (\bibinfo {year}
  {2021}),\ \Eprint{http://arxiv.org/abs/2103.16794}{arXiv:2103.16794
  [hep-ph]}%
  \bibAnnoteFile{NoStop}{Bell:2021pyy}%
\bibitem{PhysRevD.110.030001}%
  \BibitemOpen
  \bibfield{author}{%
  \bibinfo {author} {\bibfnamefont{S.}~\bibnamefont{{Navas et Al}}} (\bibinfo
  {collaboration} {Particle Data Group Collaboration}),\ }%
  \bibfield{title}{%
  \enquote{\bibinfo {title} {Review of particle physics},}\ }%
  \bibfield{journal}{%
  \Doi{10.1103/PhysRevD.110.030001}{\bibinfo {journal} {Phys. Rev. D}}\ }%
  \textbf{\bibinfo {volume} {110}},\ \bibinfo {pages} {030001} (\bibinfo
  {month} {Aug}\ \bibinfo {year} {2024}),\
  \url{https://link.aps.org/doi/10.1103/PhysRevD.110.030001}%
  \bibAnnoteFile{NoStop}{PhysRevD.110.030001}%
\end{thebibliography}%
\end{document}